\documentclass[10pt,twocolumn, compsoc]{IEEEtran}

\usepackage{array}
\usepackage[hidelinks]{hyperref}
\usepackage{wrapfig}
\usepackage{multirow}
\usepackage{color, colortbl}
\usepackage{enumitem}
\usepackage{balance}
\usepackage[skins]{tcolorbox}
\usepackage{dblfloatfix}
\usepackage{amsmath}
\usepackage{xcolor, graphicx}
\usepackage{caption}
\usepackage{subcaption}
\usepackage{adjustbox}

\newcommand{\revised}{\textcolor{black}}

\newcommand{\LESS}{\texttt{DoLesS}}

\begin{document}

\title{What Not to Test (for Cyber-Physical Systems)}

 
\author{Xiao~Ling,
        and~Tim~Menzies,~\IEEEmembership{Fellow,~IEEE}
\IEEEcompsocitemizethanks{\IEEEcompsocthanksitem X. Ling, and T. Menzies are with the Department of Computer Science, North Carolina State University, Raleigh, USA.
\protect 
E-mail: lingxiaohzsz3ban@gmail.com,  timm@ieee.org}}

\markboth{IEEE Transactions on Software Engineering}%
{Ling \MarkLowerCase{\textit{et al.}}: Faster Multi-Goal Simulation-Based Testing using {\LESS} (Domination with a Least Squares Approximation) for IEEE Journals}

\IEEEtitleabstractindextext{
\begin{abstract}

For simulation-based systems, finding a set of test cases with the least cost by exploring multiple goals is a complex task. Domain-specific optimization goals (e.g. maximize output variance) are useful for guiding the rapid selection of test cases via mutation. But evaluating the selected test cases via mutation (that can distinguish the current program from \revised{the mutated systems}) is a different goal to domain-specific optimizations. While the optimization goals can be used to guide the mutation analysis, that guidance should be viewed as a weak indicator since it can hurt the mutation effectiveness goals by focusing too much on the optimization goals.

Based on the above, this paper proposes {\LESS} ({\bf Do}mination with {\bf Le}a{\bf s}t {\bf S}quares Approximation) that selects the minimal and effective test cases by averaging over a coarse-grained grid of the information gained from multiple optimizations goals. {\LESS} applies an inverted least squares approximation approach to find a minimal set of tests that can distinguish better from worse parts of the optimization goals. When tested on multiple simulation-based systems,  {\LESS} performs as well or even better as the prior state-of-the-art, while running 80-360 times faster on average (seconds instead of hours). 

For replication purposes, all our code is on-line: \url{https://github.com/ai-se/DoLesS}.
\end{abstract}


}
\maketitle

\section{Introduction}\label{Introduction}

Simulation models play an important role in many software engineering domains. Engineers build such models to simulate complex systems~\cite{matinnejad2016automated}. In the case of cyber-physical systems, these models are sometimes shipped along with the actual device, which means that analysts can now access high-fidelity simulations of their systems. Hence, much of the work on cyber-physical testing focuses on taking full advantage of high-fidelity simulators, prior to live testing~\cite{arrieta2019pareto}.

Unfortunately, the state-of-the-art struggles with the complexities of handling multiple goals within cyber-physical systems. For example, Arrieta et al. found that, to handle N-goal problems, they had to study approximatly $N\times (N-1)/2$ pairs of two to three goals problems~\cite{arrieta2019pareto}. Hence their method took hours to terminate.

This paper introduces {\LESS}, a multi-goal mutation test method for cyber-physical systems that prunes away superfluous tests. Tests are removed if they cannot distinguish the current version of a program from something else (where ``something else'' is generated via the mutation operators described in~\S\ref{mutation}).
We will argue that:
\begin{quote}
{\em test cases from {\LESS} can detect mutants better since other methods obsess too much on \underline{{\em weak indicators}} (defined below).}
\end{quote}
To understand that, we separate system goals into $A$ and $B$:
\begin{itemize}[leftmargin=4mm]
\item
$A$ is mutation efficacy which has two sub-goals; i.e.\\ 
$A_1$: \revised{Is the selected tests optimal?} Can we select optimal tests that can mostly distinguish the current program from mutated models?\\ 
$A_2$: Can we do this task in shorter time?
\item
$B$ are the domain-specific optimization goals for the cyber-physical system. These are the domain predicates that minimize
(e.g.) processing time for images or maximize (e.g.) fuel efficiency.
\end{itemize}
{\LESS} guides its search for higher mutation efficacy (goal $A$)
by averaging over a coarse grained view of the information gained from optimizations goals $B$.
More specifically, {\LESS} does not ``dive too deep'' into the optimization information. To demonstrate the effectiveness of this approach, we explore the following research questions.

{\bf RQ1}: {\em How effective is {\LESS}, measured in terms of $A_1$?} Here we assess {\LESS} on its ability to find optimal tests that distinguish the current program from mutated models.

{\bf RQ2}: {\em How fast is {\LESS}?}. Here we assess
$A_2$ by looking at its runtime.
The {\bf RQ1} and {\bf RQ2} results will show that comparing to the prior state of the art, {\LESS} finds fewer tests which just as good (or better) at distinguishing between the current program and mutated models.
Further, it can run 80-360 times faster across 20 trials. This means the task that previously took hours to complete can be completed in less than a minute with {\LESS}.

After that, we move to more surprising results\revised{:}

{\bf RQ3:} {\em Does maximizing for goal $A$ (mutation effectiveness) mean compromising on optimization goals $B$?} 
\revised{At first glance, the values on the optimization goals seem to tell a negative story about {\LESS} (lower optimization goals than other optimizers).
However, it turns our that the relationship between mutation effectiveness and optimization goals has some
{\em previously unreported non-linear effects}. Specifically, if we are over-zealous in
exploring the optimization goals, it may not actually improve the efficacy goals.  
In those cases where {\LESS} achieved supposedly-worse results (on optimization goals), they {\em only appear in the mutated systems}. That is, we only observe {\LESS} being inferior in {\em corrupted mutations}.
Also, the good news here is that {\LESS} actually can achieve better scores on mutation efficacy (goal $A$) than other methods. Thus, we say ``yes'' to this RQ that maximizing for the mutation effectiveness means relaxing on goals $B$. }

{\bf RQ4:}  {\em Should mutation testing tools for cyber-physical systems emphasis goals $A$ or $B$?}
We will show (in Figure~\ref{fig:discussion2}) that the relationship between goals $A$ and $B$ is not linear. Up until some point, chasing $B$ does improve mutation efficacy (goal $A$). However, after some turning point, further pursuit of optimization goals will not improve the  mutation effectiveness too much. Hence our answer to {\bf RQ4} is that, 
if the goal is mutation effectiveness, then we need to relax on pursuing the optimization goals within the corrupted mutation (since it is not very informative to optimize the wrong program).

If readers are surprised by the {\bf RQ3} and {\bf RQ4} results, we hasten to point out that something like this result has been seen before (in machine learning). A repeated result~\cite{ratner2019training,Pornprasit23,wang2019characterizing,nair2017using} is that when optimizing for some goals $A$, it is possible to be guided by some easy-to-compute values $B$, even if $B$ is only a weak indicator for $A$~\cite{ratner2019training}:
\begin{itemize}[leftmargin=4mm]
\item
Some researchers~\cite{Pornprasit23} use generative models to augment supervised learning. While the results from the generative model may not in highest quality, all these outputs provide hints on how to better direct another algorithm (e.g. a machine learner).
\item
Researchers explore test case selection via crowd-sourcing since this is a fast way to collect many diverse opinions. While the values of opinions (for achieving $A$) might be questionable, they can be useful in the aggregation~\cite{wang2019characterizing}.  
\item  
Effective automatic configuration tools can be built from regression trees with just a few examples. Even if those regression trees are very poor predictors of performance (in an absolute sense), they can still be useful to rank (in a relative sense) different configurations.
In this way, Nair et al. were able to find the one percent best configurations using regressions trees with error rages as high as
90\%~\cite{nair2017using}.  
\end{itemize}
The lesson of this paper is that
for simulation-based systems, finding  test cases (with   least effort) by exploring multiple goals is a complex task. 
Domain-specific optimization goals (e.g. maximize output variance) are useful for guiding the rapid selection of test cases via mutation. That said:
\begin{itemize}[leftmargin=4mm]
\item
Selecting test cases via mutation (that can distinguish the current program from \revised{the mutated systems}) is a {\em different goal} to domain-specific optimizations.
\item
While the optimization goals can be used to guide the mutation analysis, that guidance should be viewed as a weak indicator since it can hurt the mutation efficacy due to focusing too much on the optimization goals.
\end{itemize}

\noindent
In summary, we say our novel contributions are:
\begin{enumerate}[leftmargin=6mm]
\item
We verify that test case selection problem is not trivial. More specifically, this problem cannot be solved just by applying standard optimizers (NSGA-II, NSGA-III and MOEA/D) in off-the-shelf manner. Due to the many-goal nature of the problem, we need somehow to extend our optimizing technology\revised{.}\footnote{In the literature, ``many-goal'' usually refers to 2 or 3 goals and ``multi-goal'' refers to four or more.}
\item 
We propose a novel test selection method {\LESS}.
\item
We verify and explain the rationale on why off-the-shelf multi-goal optimizers failed on this task by conducting additional experiment and analysis.
\item
We clearly document the value of doing {\LESS}. When testing on six cyber-physical models, {\LESS} finds test suites as good, or even better, than those found by Arrieta et al.'s approach~\cite{arrieta2019pareto}. Further, {\LESS} does so while running 80-360 times faster (seconds instead of hours, mean time). Hence, we recommend {\LESS} as a fast method to find minimal test cases for multi-goal cyber-physical systems.
\end{enumerate}
The rest of this paper is structured as follows. Section~\ref{background} introduces the background and related work in test case selection for simulation-based testing. Section~\ref{methodology} introduces the problem of studying effectiveness measurement metrics in cyber-physical systems and illustrates how they are calculated by mathematical formula. Moreover, multi-objective optimizers and our proposed approach are introduced in this section as well. Section~\ref{experiment} introduces 
the case studies, performance evaluation metrics, and statistical analysis method used in this study. Section~\ref{result} shows our experimental results. Section~\ref{discussion} discusses the generality of {\LESS}, Section~\ref{threats_to_validity} explores threats to validity and Section~\ref{conclusion} makes the summary of our study and states the possible future work.


\section{Background and Related Work}\label{background}
A repeated result is that test suites can be {\em minimized} (i.e. we can run fewer tests) while still being as {\em effective} (or better) than running the larger test suite~\cite{ahmed2016test, di2015coverage, wong1997test, wong1998effect, yoo2012regression}. Note that ``effective'' can mean different things in different domains, depending on the goals of the testing. For example, at FSE'14, Elbaum et al.~\cite{elbaum2014techniques} reported that Google could find similar number of bugs, but after far fewer tests execution. This was an important result since, at that time, the initial Google test suites were taking weeks to execute. Such long test suite runtimes is detrimental to many agile software practices.

\begin{figure*}[!b] 
    \centering
    \includegraphics[width=0.75\textwidth]{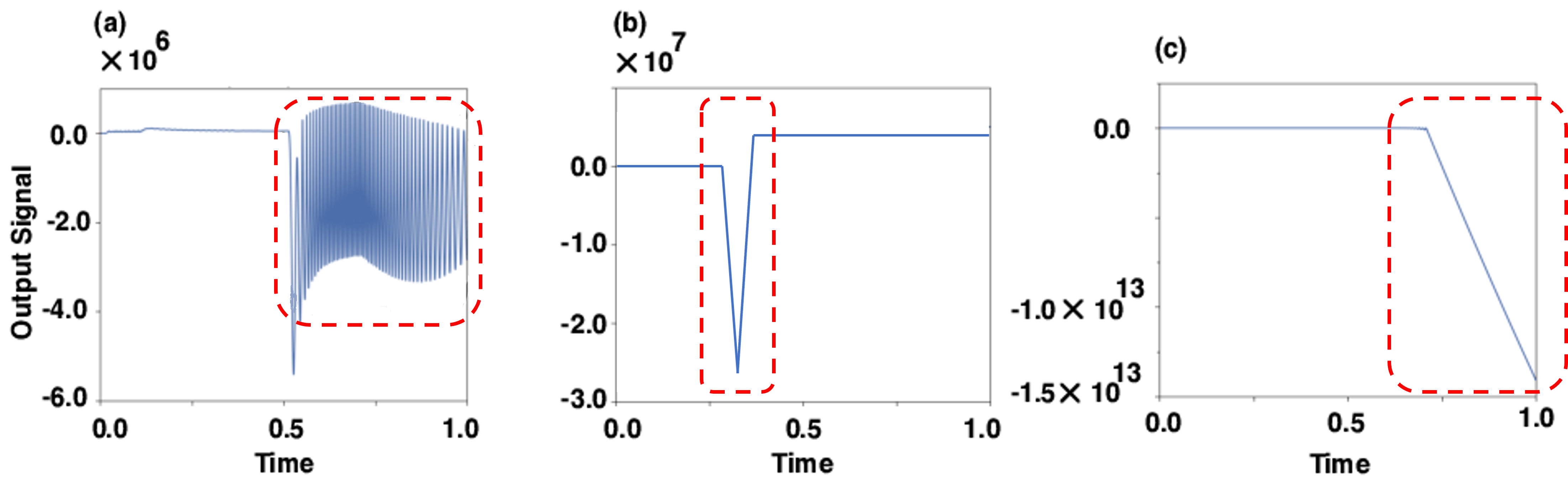}
    \caption{Examples of anti-patterns seen for systems under feedback. The name of anti-patterns from left to right are {\em instability}, {\em discontinuity}, and {\em growth to negative  infinity} correspondingly. The alarming patterns are shown in red marks.}
    \label{fig:antiPattern}
\end{figure*}

{\it Test case selection for traditional software testing:} Research has found many test case selection techniques such as DejaVu based, firewall based, dependency based, and specification based techniques~\cite{engstrom2010systematic}. We note that different test suite minimization methods need different kinds of data. For example, in 1995, Binkley et al. proposed a semantic-based method which takes the use of differences and similarities of two consecutive versions to select test cases~\cite{binkley1995reducing}. Rothermel et al. developed a test case selection technique for C++ software on 2020~\cite{rothermel2000regression}. In 2001, Chen et al. developed test case selection strategies based on the boolean specifications~\cite{chen2001test}. In 2005, a fuzzy expert system was developed in test case selection by Xu et al.~\cite{xu2005application}. In 2006, Grindal et al. presented an empirical study on evaluating five combination strategies for test case selection~\cite{grindal2006evaluation}. In 2007 to 2015, multi-objective search genetic algorithms got more attention from software testing researcher. For example, In 2014, Panichella et al. introduced MOGA which increases diversity by injecting new orthogonal individuals during test selection search process based on the mechanisms of orthogonal design and orthogonal evolution~\cite{panichella2014improving}. In 2015, Mondal et al.~\cite{mondal2015exploring} firstly compared ``test case diversity'' and ``code coverage'' in several real-world case studies and then proposed new test selection technique based on NSGA-II which maximize both two criteria. Also, in 2015, Souza et al.~\cite{de2015hybrid} proposed two MOO algorithms and present the empirical evaluation of their performances in test selection compared to NSGA-II and MBHS. In 2017, Devroey et al.~\cite{devroey2016search} formulated the test case selection problem in product line from a Feature Transition System (FTS) as an optimization problem which maximizes the coverage measure in FTS while minimizes/maximizes the number of products needed to execute all test cases. Also in 2017, an extensive study was made by Devroey et al.~\cite{devroey2017dissimilar} which diverse a distance function between actions of FTS and products on which the test case may be executed.

{\it Test case selection for model-based  software}: Another large group of model in software testing is the simulation-based software. In 2010, Hemmati et al.~\cite{hemmati2010industrial} explores multiple similarity measurements to support the similarity testing in State Machine models. Moreover, one year later, Cartaxo et al. implemented a similarity function for test case selection in model-based testing~\cite{cartaxo2011use}. Pradhan et al. ~\cite{pradhan2016search} proposed a multi-objective optimization test case selection approach which can be used with limited time constraints. Arrieta et al. also used test case execution history to select test cases~\cite{arrieta2016search}. Same in 2016, Arrieta et al. also conducted a study on test case selection of cyber-physical product line where their proposed method can be adopted to "X-in-the-Loop" test level~\cite{arrieta2016search1}. In 2017, Lachmann et al.~\cite{lachmann2017multi} did an empirical study on several black-box metrics and made comparisons on their performance in selecting test cases in system testing. On 2020, Fremont et al. conducted a test selection study which they presented a new approach to automated scenario-based testing of the safety of autonomous vehicles~\cite{fremont2020formal}.

Due to the obtainable data in simulation models, many of the above methods are unsuitable for cyber-physical systems, for two reasons.
{\em Firstly}, cyber-physical systems are embodied in their environment. Hence, it is not enough to explore static features of (e.g.) the code base. Rather, it is required to test how that code base reacts to its surrounding environments. Hence, using just static information such as (e.g.) code coverage metrics is not recommended for testing cyber-physical systems. 

~
{\em Secondly}, at least for the systems studied here, cyber-physical systems make extensive use of process control theory. In that theory, the feedback controller is used to compare the value or status of process variables with the desired set-point. This controller then applies the difference as a control signal to bring the process variable output of the plant to the same value as the set-point. Hence, for test suite minimization of process control applications, the requirement is data collected from the feedback loops inside the cyber-physical systems. Accordingly, here we use input and output signals in the simulation models instead of execution history or coverage information.

In one of the IST'19 journal paper, Arrieta et al.~\cite{arrieta2019pareto} explored issues associated with test suite minimization by using the data extracted from feedback loops. They noted that feedback loops have anti-patterns; i.e. undesirable features that appear in a time series trace of the output of the system. The reason why anti-patterns are important factors to be considered in test minimization for simulation models is because a faulty behavior in simulation model may not cause the execution interruption as traditional software does. For example, if a correct {\bf $+$} operator is mistakenly coded to {\bf $-$}, the whole system can still run successfully. For another example, the model can still run with no error if the {\bf OR} logic operator is replaced to {\bf AND}. Thus, traditional scenarios such as execution failing history has very limited impact to test case minimization in simulation models. In such situation, anti-patterns are much more important since they can directly reveal undesirable behaviors from the output signals even the model is executed successfully. Figure~\ref{fig:antiPattern} shows three such features correspondingly from left to right in the red dash rectangle: {\em instability}, {\em discontinuity}, and {\em growth to infinity}. Later in this paper we will mathematically define these anti-patterns.

In all, Arrieta et al.~\cite{arrieta2019pareto} explored seven goals for cyber-physical model testing: maximizing the three anti-patterns that shown in Figure~\ref{fig:antiPattern}, maximizing three other measures of effectiveness, as well as minimizing total execution time. Arrieta et al. used {\em mutation testing} to check the validity of their minimized test suite. Mutation based testing is a fault-based testing technique which implements ``mutation adequacy score'' to assess test suite adequacy by creating mutants~\cite{jia2010analysis}, and then pruning test cases which cannot distinguish the original model from the mutant. In our study, we use the mutants generated in the experiments from Arrieta et al.~\cite{arrieta2019pareto}. Those mutants were generated with Hanh et al.'s technique~\cite{binh2016novel} and some of the mutants are removed if (a) they are not detected by any test case, (b) they are killed by all test cases, and (c) they are equivalent mutants~\cite{papadakis2015trivial}. Like Arrieta et al., we say a test suite is {\em minimal} when it retires as many mutants as a larger suite.

Mutation testing is the inner loop of Arrieta et al.'s process and, in their experiments, they found mutation testing to be an effective technique. The problem area in their work was the outer loop that optimized for seven goals. They found that standard optimizers such as NSGA-II~\cite{deb2002fast} can be ineffective for more than three goals (a result that is echoed by prior work~\cite{sayyad2013value}). More recent optimizers like NSGA-III~\cite{deb2013evolutionary} and MOEA/D~\cite{zhang2007moea} also failed for this multi-goal task. Later in this paper, we replicate their experiment and strengthen that finding (see RQ1).

To address this optimization failure, they resorted to ``pairwise'' approach based on NSGA-II. That is, they ran NSGA-II with all 21 subsets of ``choose two or three from seven'' goals then returned the test suite associated with the run which has the best scores (where the ``best'' here is measured just on a subset of goals). 
While their work is definitely an extension of state-of-the-art, their study has two drawbacks. Firstly, the test cases selected in this way is only the best which measured on a subset of the optimization goals. Secondly, the ``pairwise'' approach increases optimization time by an order of magnitude, which is a major issue for large simulators, especially when we are running these algorithms 20 times (to check the generalizability of this stochastic process).

\begin{figure}[t]
    \centering
    \includegraphics[width=0.48\textwidth]{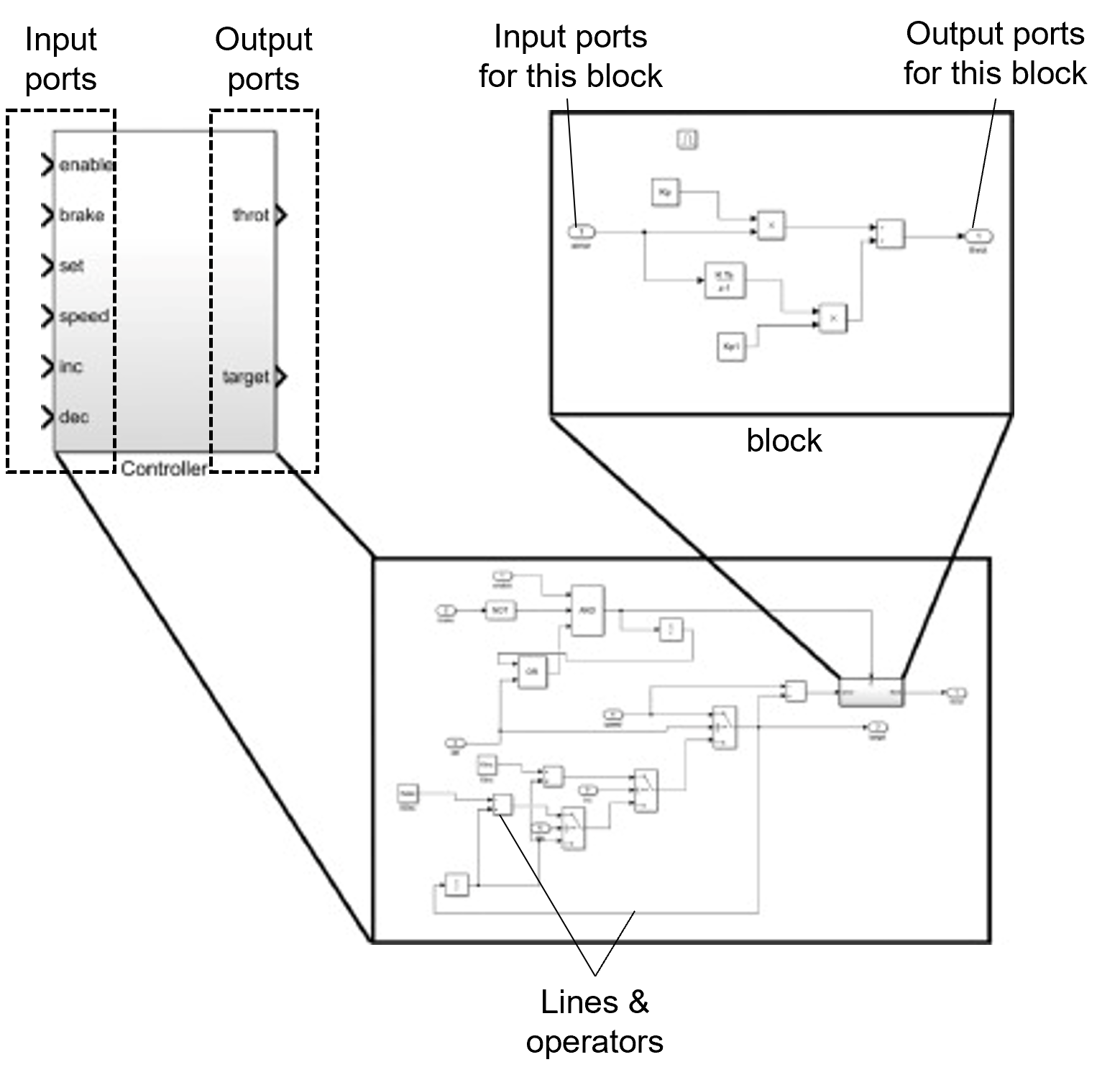}
    \caption{Simple example of a Simulink model - Cruise Controller of a car~\cite{arrieta2019pareto}. }
    \label{fig:simulink}
\end{figure}

\begin{figure*}
    \centering
    \includegraphics{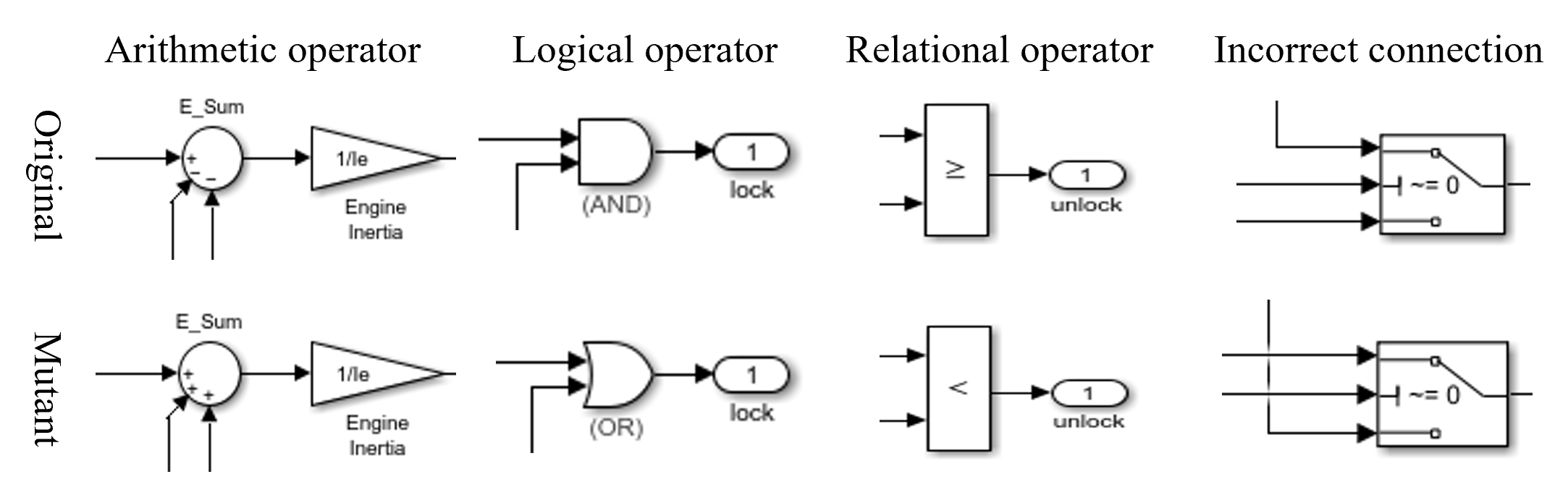}
    \caption{Examples of mutants in Simulink models. From left to right, arithmetic mutant switch the arithmetic operators; logical mutant switch the logical operators; relational mutant change the relational operators to opposite site; and connection mutant switch the input lines of the ``Switch'' block.}
    \label{fig:example_mutant}
\end{figure*}

Hence in this work, we seek to improve the mutation based test suite minimization method from Arrieta et al.'s study~\cite{arrieta2019pareto}. Like them, we will optimize for the anti-patterns and effectiveness measures seen in process control systems. But unlike that prior work, we will offer methods that simultaneously succeed across many goals (without needing anything like the pairwise heuristic used in Arrieta et al.). Further, we show that all this can be achieved without   additional runtime cost.

\subsection{Testing Simulation Models}\label{simulationModel}
Cyber-physical system developers often use simulation tool (e.g. Simulink) to build cyber-physical models~\cite{chowdhury2018automatically}. For an example Simulink models, see Figure~\ref{fig:simulink}. This is a model with two hierarchical levels~\cite{arrieta2019pareto}. A complex model will have far more blocks and operators.

In Simulink models, the inputs and outputs are all signals (here signal means a time series function). This means at each simulation time step $\Delta T$, there will be a value in each input and in each output regarding to that time step. For example, if we simulate a model for 5 seconds in real time and the time step $\Delta T$ is 0.05, then there will be $5/0.05 + 1 = 101$ simulation steps, which means each input or output should be a vector of length 101.

Since the inputs of simulation models are vectors of time series, a test case $t$ of the simulation model will also be $n$ time series vectors $t = \{ t_{i_1}, t_{i_2}, \cdots, t_{i_n} \}$ where $n$ is the number of inputs.
Assuming an initial set of $m$ test cases $\{t_1, \cdots, t_m\}$ for a simulation, each test simulates the model from a set of unique $k$ input signals $\{ is_1, \cdots, is_k \}$ to a set of $l$ output signals $\{ os_1, \cdots, os_l \}$~\cite{matinnejad2016automated, arrieta2019pareto}.

Our goal for this study is to select representative test cases from the initial test suite to minimize the test execution time, but not influence the testing performance. Here we can define the test case selection problem as follow:
\begin{tcolorbox}[boxsep=1pt,left=4pt,right=4pt,top=2pt,bottom=2pt]\small
    Given an initial test suite $T$ which includes $n$ test cases $T = \{ t_1, t_2, t_3, \cdots, t_n \}$ and an evaluation function $f$ which can evaluate the fault detection ability of a test suite, the goal of the test case selection is to find a new test suite $TS = \{ ts_1, ts_2, \cdots, ts_a \}$ such as $TS \subseteq T$ and $f(TS) = f(T)$.
\end{tcolorbox}
If we search in the space which contains all the subsets of the initial test suite, then the search space will be very large. For example, with only 100 test cases, there will be $2^{100}-1$ possible subsets. Thus, cost-effectively selecting test cases is a significant problem.

There are two key factors in test case selection for simulation models. {\em Simulation runtime} indicates the total time to simulate test cases for the necessary information (e.g. coverage, test execution history, and effectiveness measurements used in this study). When simulation models explore high fidelity systems (such as electronic models, physical models, AV, and drone simulation models), they can be slow to execute~\cite{arrieta2016search}. For example, previous literature reported that testing a high fidelity simulation model can take hours to days~\cite{arrieta2019pareto, sagardui2017multiplex,gonzalez2018enabling}. This factor mostly is \textit{not decreaseable} because all test case selection algorithms require test case information from simulation, and such simulation time is not changeable.

On the other hand, {\em algorithm runtime}  
indicates the time for test case selection algorithm. This factor \textit{can} be minimized by finding a fast approach (e.g. in our study, we find {\LESS} runs significantly faster to find minimum test suite). In this study, we try to reduce the difficulty of test case minimization by finding minimum test suite in a very short feedback loop (i.e. fast algorithm runtime).

\subsection{Mutation Testing in Simulation Models}\label{mutation}

One persistent problem in SE testing is the ``oracle problem''; i.e where can we get the expertise that can judge if a test suite is ``good enough''. The insight offered by mutation testing~\cite{offutt1996experimental, schuler2009javalanche, papadakis2019mutation, pizzoleto2019systematic, jia2010analysis} is that it is possible to automate the generation of a test oracle as follows:
\begin{itemize}[leftmargin=4mm]
\item Suppose we have a program $P$ and a test suite $T$.
\item That program $P$ can be ``mutated'' to generate a syntactically valid piece of code, with some changes (e.g. swap a minus sign with a plus or any other mutations listed in Table~\ref{tab:fault_pattern}).
\item The least we should expect from $T$ is that it  can recognize when $P$'s behavior change (since otherwise, the tests are blind to operational changes in the code).
\item Given a simulation program $P$ mutated to $P'$, we say that a test suite is ``good enough'' if measurements in the output signal taken from $P$ and $P'$ are different \footnote{Our pre-experimental expectation was that  statistical methods would be required to detect nuanced changes between  $P$ and $P'$. But, referring to Figure~\ref{fig:antiPattern}, the faults induced but our mutants tend towards very large changes in those measurements. Hence, just computing residuals sufficed for this recognizing differences. }.
\end{itemize}
In traditional software engineering projects, mutants are systematically seeded by changing a piece of code from the original project. However, mutants are quite different in the simulation models. As Figure~\ref{fig:simulink} shows, simulation models are combined with multiple blocks, lines, and operators instead of pure lines of code. Therefore, mutating simulation models (e.g. Simulink projects) requires certain fault patterns~\cite{matinnejad2018test}. Using a literature review, we found numerous typical fault patterns for the simulation models (see Table~\ref{tab:fault_pattern}). 
Moreover, we list some examples of basic mutants from Simulink models in Figure~\ref{fig:example_mutant}. In that figure, (a) the mutant of arithmetic operator changes the operation signs (e.g. + to -); (b) the mutant of logical operator switches the logic sign (e.g. AND to OR); (c) the mutant of relational operator changes the relational sign (e.g. $>=$ to $<$; and (d) the mutant of the connection block switches its input lines. As introduced in Section~\S\ref{simulationModel}, the outputs of simulation models are time series vectors. Therefore, we say that a test case detects the mutant if the output vector from mutated model has some different values comparing with the output vector from original model (e.g. mutated model has output signal $[0.1, 0.2, 0.3, 0.5, 0.3, 0.1]$ and original model has output signal $[0.1, -0.3, -0.5, -0.5, 0.3, 0.1]$).\\
As a sanity check, we manually inspect the mutants generated outputs from  Arrieta et al.~\cite{arrieta2019pareto} before running our experiments to check if the generated mutants follow these fault patterns. The inspection result indicates that all observed mutants are satisfy one of the fault patterns shown in Table~\ref{tab:fault_pattern}.

\begin{table}[t]
    \centering
    \caption{Simulation fault patterns collected from literature~\cite{binh2012mutation, brillout2009mutation, yin2014research, zhan2005search}, and also listed in~\cite{matinnejad2018test}.}
    \label{tab:fault_pattern}
    \begin{tabular}{p{0.15\textwidth}p{0.25\textwidth}}
        Pattern & Illustration \\
        \hline
        Change of constant values & Change constant value $c$ to some other values if it is numeric. If it is boolean, negating it.\\
        \hline
        Change of arithmetic operators & Switch +/- or replace + to $\times$. \\
        \hline
        Change of relation operator & Switch $\geq$ to $<$ or $\leq$, and switch $\leq$ to $>$ or $\geq$. \\
        \hline
        Change of logical operator & Switch \textbf{AND}, \textbf{OR} and \textbf{XOR}; Remove \textbf{NOT} or add \textbf{NOT}. \\
        \hline
        Incorrect Connection & Switch the input lines of the ``Switch'' block. \\
        \hline
        Incorrect Signal Data Type & Switch the ``double'' data type to ``single'', or switch ``fixdt(0,8,3)'' data type to ``fixdt(0,8,2)''. \\
        \hline
        Wrong Initial conditions and delay values & Change the initial value in ``Integration'' and ``Unit Delay'' blocks.
    \end{tabular}
\end{table}

\section{Experimental Methods}\label{methodology}
\subsection{Simulation Effectiveness Metrics}\label{blackboxData}
In this study, we implement five out of seven effectiveness measurement metrics which Arrieta et al.~\cite{arrieta2019pareto} used in their study. The first three metrics are also widely used in previous studies~\cite{wang2013minimizing, matinnejad2015effective, matinnejad2017automated}, and the forth metric is proposed by Arrieta et al.~\cite{arrieta2019pareto}. Moreover, since the values in every metric differ a lot, we normalized all metrics before performing search on them to avoid the threat from different scales.


Please note that in all following 5 metrics, $T$ represents the initial test suite such that $T = \{t_1, t_2, \cdots, t_n\}$ and $n$ represents the number of tests in the initial test suite. If test case $i$ is selected, then $Select_{t_i} = 1$. Otherwise, $Select_{t_i} = 0$. Moreover, $TS$ in the following formulation means the new test suite after the selection process.

Note that since all our case studies are mined from public resources, they are well-developed and may not have enough records on faulty behavior. To satisfy our experiment purpose, we utilized mutation testing and generated as much mutants as possible by following the Simulink faulty patterns introduced in Section~\S\ref{mutation} to reproduce the faults that may occur in the product line. Hence all those metrics are collected from the mutated environment instead of the original environment.

\subsubsection{Test Execution Time}
Total test execution time is the first metric we implement in our study. Wang et al.~\cite{wang2013minimizing} stated that the number of selected test cases can be treated as the measurement for selecting representative test cases from the initial test suite. However, in simulation models, different test case may have different execution time. Thus, we cannot directly count the number of selected test cases in our experiment. To mitigate that, we adopt the idea from Arrieta et al.~\cite{arrieta2019pareto} that using the overall execution time of the selected test cases as the search guidance.~\\~
To say more specifically, let $tet_i$ be the the normalized test execution time for the test case $i$, the overall test execution time for a test suite can be calculated as follow
\begin{equation}
    executionTime(TS) = \sum_{i=1}^{n} Select_{t_i} * tet_i
\end{equation}
In our study, we want to \underline{{\em minimize}} this metric because the goal of test case selection is to decrease the test execution time.

\subsubsection{Discontinuity in Output Signal}
Discontinuity is the second metric we implement in our study. As Matinnejad et al.~\cite{matinnejad2017automated} stated, the discontinuity of the output signal is a short duration pulse in the output signal, which means the output signal increases or decreases to a value in a very short time, and recovers back to normal. If executing a test case causes discontinuity in the output signal, then that test case detects the faulty behavior in the model.
To be more specific, the discontinuity score $discontinuity(O_j)$ of an output signal $j$ can be calculated as follow:
\begin{equation}
    discontinuity(O_j) = \max\limits_{dt=1}^{3}(\max\limits_{i=dt}^{k-dt}(min(lc_i, rc_i)))
\end{equation}
where 
\begin{itemize}[leftmargin=4mm]
    \item $lc_i = |sig(i \cdot \Delta t) - sig((i-dt) \cdot \Delta t)| / \Delta t$ is the left change rate of step $i$. $\Delta t$ is the time stamp.
    \item $rc_i = |sig((i+dt) \cdot \Delta t) - sig(i \cdot \Delta t)| / \Delta t$ is the right change rate of step $i$.
\end{itemize}

The discontinuity score of a test case will be the sum of the discontinuity score for each output signal. Assume the normalized discontinuity score for the test case $i$ is $dc_i$, the overall discontinuity score for a test suite can be calculated as follow
\begin{equation}\label{tscalculation}
    discontinuity(TS) = \sum_{i=1}^{n} Select_{t_i} * dc_i
\end{equation}

In our study, we want to \underline{{\em maximize}} this metric because the goal is to detect more discontinuity in the output signal.

\subsubsection{Instability in Output Signal}
Instability is the third metric we implement in our study. As Matinnejad et al.~\cite{matinnejad2017automated} stated, the instability of the output signal is a duration of quick and frequent oscillations in the output signal, which means the output signal increase and decrease repeatedly in a duration of time. If executing a test case causes instability in the output signal, then that test case detects the undesirable impact on physical process~\cite{matinnejad2017automated}. 
To be more specific, the instability score $instability(O_j)$ of an output signal $j$ can be calculated as follow:
\begin{equation}
    instability(O_j) = \sum_{i=1}^{k}|sig(i \cdot \Delta t) - sig((i-1) \cdot \Delta t)|
\end{equation}
where $k$ is the total number of simulation steps and $\Delta t$ is the time stamp in the simulation model for each step.

Similar to the discontinuity metric, the instability score of a test case will be the sum of the instability score for each output signal. The overall instability score will be calculated similar to the Equation~\ref{tscalculation}. For the space reason, we are not listing it again here.

In our study, we want to \underline{{\em maximize}} this metric because the goal is to detect more instability in the output signal.

\subsubsection{Growth to Infinity in Output Signal}
This is the forth metric we implemented in our study. As Matinnejad et al.~\cite{matinnejad2015effective} pointed out, the growth to infinity of the output signal is the phenomenon that the output signal increases or decreases to infinity value. If executing a test case causes growth to infinity in the output signal, then that test case detects the faulty behavior in the model. 
To be more specific, the growth to infinity score $infinity(O_j)$ of an output signal $j$ can be calculated as follow:
\begin{equation}
    infinity(O_j) = \max\limits_{i=1}^{k} |sig(i \cdot \Delta t)|
\end{equation}
where $k$ is the total number of simulation steps and $\Delta t$ is the time stamp in the simulation model for each step.

Same as above, the infinity score of a test case will be the sum of the infinity score for each output signal. Moreover, the overall infinity score of a test suite will be the sum of infinity score of selected test cases (e.g. Equation~\ref{tscalculation}).

In our study, we want to \underline{{\em maximize}} this metric because the goal is to detect more growth to infinity situation in the output signal.

\subsubsection{Output Minimum and Maximum Difference in Output Signal}
This is the last metric we implemented in our study. Arrieta et al. proposed this metric in their work because the difference between maximum output signal and minimum output signal can indicate the level of how a model is being tested. If executing a test case results in large minimum and maximum difference in the output signal, then that test case can detect more parts in the simulation model. 
To be more specific, the minmax score $minmax(O_j)$ of an output signal $j$ can be calculated as follow:
\begin{equation}
    minmax(O_j) = |\max\limits_{i=1}^{k}(sig(i \cdot \Delta t)) - \min\limits_{i=1}^{k}(sig(i \cdot \Delta t))|
\end{equation}
where $k$ is the total number of simulation steps and $\Delta t$ is the time stamp in the simulation model for each step.

Same to the above metrics, the minmax score of a certain test case will be the sum of minmax scores for all output signals. Moreover, the overall minmax score for a test suite will be calculated like Equation~\ref{tscalculation}.

In our study, we want to \underline{{\em maximize}} this metric because the goal is to coverage more parts that can be tested.

\begin{figure*}[t]
    \centering
    \includegraphics[width=\textwidth]{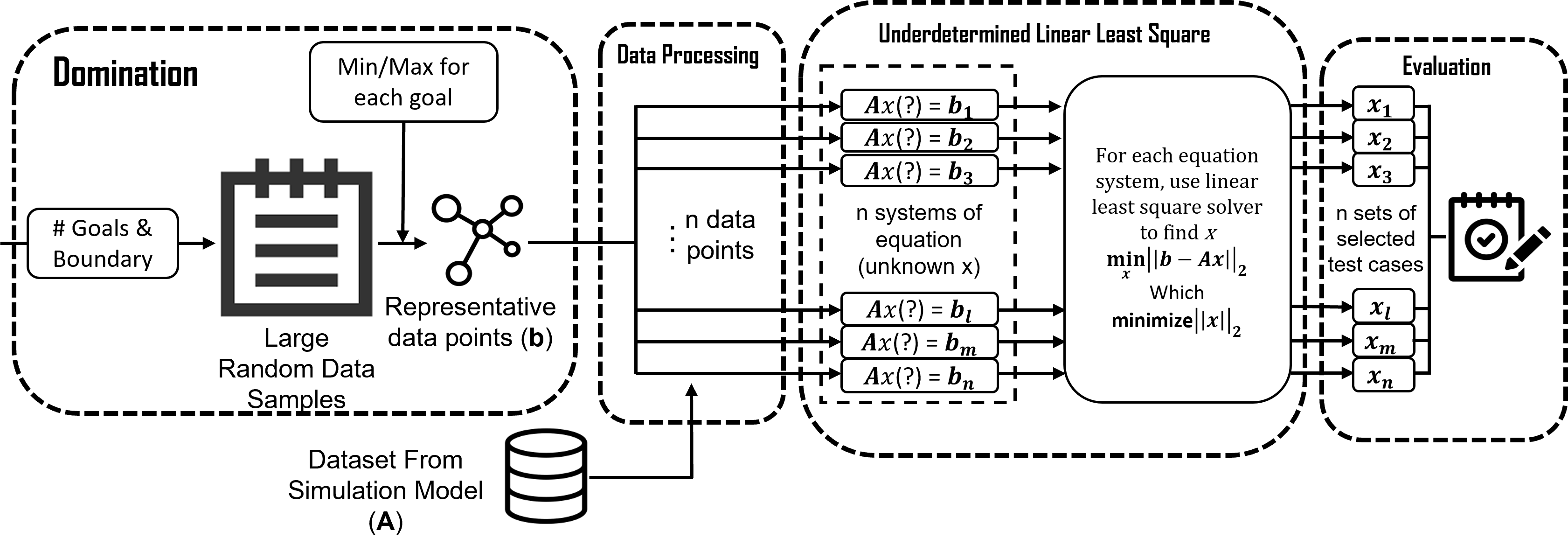}
    \caption{{\LESS} Framework. The entire framework is consisted by \textit{domination block}, \textit{data processing block}, \textit{linear least square block}, and \textit{evaluation block}}.
    \label{fig:System Framework}
\end{figure*}

\subsection{Algorithms}\label{algorithm}
\subsubsection{Binary vs Continuous Domination}
In the following, all the algorithms use {\em binary
domination} except for {\LESS} that uses {\em continuous domination}.

{\em Binary domination} say one individual is better than another if it is better on at least one goal and worse on none. Many studies~\cite{Zitzler2004,sayyad2013value,Wagner07} warn that binary domination struggles to distinguish candidates for three or more goals.

For many-goal problems, Zilter's {\em continuous domination} predicate~\cite{Zitzler2004} is useful~\cite{Zitzler2004,sayyad2013value,Wagner07}. Continuous domination judges the domination status of pair of individuals by running a ``what-if'' query which checks the situation when we jump from one sample to another, and back again. Specifically:
\begin{itemize}
    \item For the forward jump, we compute \\
    $s_1 = -\sum_i  e^{w_i * (a_i-b_i)/n}$.
    \item For the reverse jump we compute \\
    $s_2 = -\sum_i  e^{w_i * (b_i-a_i)/n}$.
\end{itemize}
where $a_i$ and $b_i$ are the values on the same index from two individuals, $n$ is the number of goals (in our case $n=5$), and $w_i$ is the weight \{-1,1\} if we are minimizing or maximizing the goal $i$ correspondingly. According to Zitler~\cite{Zitzler2004}, one example is preferred to another if we lost the least jumping to it; i.e. $s_1<s_2$.

Specifically, in this work, we use this predicate to select better goal sets that (a) {\bf minimize} test execution time, (b) {\bf maximize} discontinuity score, (c) {\bf maximize} instability score, (d) {\bf maximize} growth to infinity score, and (e) {\bf maximize} output minimum \& maximum difference.

\subsubsection{NSGA-II} NSGA-II is a common evolutionary genetic algorithm~\cite{deb2002fast}. Firstly, it generates an initial set of population as the starter of the entire algorithm. Secondly, these candidates will evolve to offsprings in a series of generations by implementing the crossover and mutation operators with their individual probability. In our reproduction experiment, we use single point crossover with 0.8 probability and bit-flip mutation with $1/N$ probability ($N$ is the number of test cases). Thirdly, parents for next generation will be selected by selection operator, which utilizes a non-dominated sorting algorithm to select top non-dominated solutions~\cite{panichella2017automated}. In the situation where a front needs to be divided because it exceeds the total number of population, NSGA-II uses the crowding distance to split candidates in that group.

\subsubsection{NSGA-III} NSGA-III is an improved NSGA-II algorithm~\cite{deb2013evolutionary}. In NSGA-III, all procedures such as initial population generation, crossover, and mutation are similar to NSGA-II, except selection procedure. In NSGA-III, the selection procedure is applied based on a set of reference points. The reference points are uniformly distributed on the normalized hyper-plane with some division number $p$~\cite{deb2013evolutionary}. After that, each objective point is normalized adaptively and associated with a reference point by calculating its distance to the corresponding reference line. Niche-Preservation operation is then applied to select candidates which will be used in the next generation~\cite{deb2013evolutionary}.

\subsubsection{MOEA/D} MOEA/D~\cite{zhang2007moea}   decomposes the problem into multiple sub-problems with less objectives in each subgroup and solves these sub-problems simultaneously~\cite{zhang2007moea}. To do so, prior to inference, all examples get random weights assigned to their goals. Examples are then clustered by those weights such that all examples know the space of other examples that weighted in a similar direction. Next, during the execution, if one example $X_0$ finds a way to improve itself, its local neighborhood will move in the same direction as $X_0$.

\subsubsection{{\LESS}}\label{less}
Figure~\ref{fig:System Framework} shows the entire framework of our approach. Unlike above evolutionary algorithms, our proposed approach {\LESS} ({\bf Do}mination with {\bf Le}a{\bf s}t {\bf S}quares Approximation) does the optimization from the left most block to the right most block in one traverse. More specifically, there are four main blocks in our framework:

{\em Continuous domination block} (defined below, see the first block in Figure~\ref{fig:System Framework}) is used to reduce the size of initial large random sets of goals and find a ``best'' group of representative samples. 
After sorting via domination, {\LESS} divides data into:
\begin{itemize}[leftmargin=4mm]
    \item The $\sqrt{n}$ ``best'' items. We randomly generate 10000 initial candidates so the first 100 candidates with highest domination score are grouped   the ``best'' group.
    \item And the remaining ``rest'' items.
\end{itemize}
Here we set final population to 100 since: (a) 10000 random initial population is large enough to cover a wide range of possible outcomes and (b) to make comparison fair, we select same number of final candidates as previous work~\cite{arrieta2019pareto}.

In the {\em data processing block} in Figure~\ref{fig:System Framework},
we take data from Simulink models, and process it into the form of least square approximation structure by combining with the representative goals generated from {\em continuous domination}. 
Table~\ref{tab:exampleBB&TC}(i) shows a simple example of effectiveness measurement data collected from the models. We can find that each test case will have a single score for all effectiveness measurement data ($a_{ij}$ means the score of test case $i$ in effectiveness measure $j$). The corresponding matrix equation system for the above example is shown in Table~\ref{tab:exampleBB&TC}(ii). This equation shows the linear relationship of test selection outcomes and the final effectiveness measure scores (e.g. the final score of effectiveness measure 1 can be obtained by $em_1 = a_{11} \cdot t_1 + a_{12} \cdot t_2 + a_{13} \cdot t_3 + a_{14} \cdot t_4$ where $t_i$ is the outcome of test selection). In this example, our goal is to find the best outcomes of $t_1$ to $t_4$ which can result $em_1$ to $em_5$. To summarize the above example, in our approach, we collect effectiveness measurement data for $n$ test cases (like Table~\ref{tab:exampleBB&TC}(i)) and want to find the best set of outcomes for $t_1$ to $t_n$ which can get the closest scores to representative goals which are selected by continuous domination.

\begin{table}[t]
    \small
    \centering
    \caption{An example of (i) collected effectiveness measurement data (EM means effectiveness measure) and (ii) its corresponding matrix equation form}
    \begin{tabular}{c}
        \begin{tabular}{c|c|c|c|c}
             & $t_1$ & $t_2$ & $t_3$ & $t_4$ \\
            \hline
            EM 1 & $a_{11}$ & $a_{12}$ & $a_{13}$ & $a_{14}$ \\
            \hline
            EM 2 & $a_{21}$ & $a_{22}$ & $a_{23}$ & $a_{24}$ \\
            \hline
            EM 3 & $a_{31}$ & $a_{32}$ & $a_{33}$ & $a_{34}$ \\
            \hline
            EM 4 & $a_{41}$ & $a_{42}$ & $a_{43}$ & $a_{44}$ \\
            \hline
            EM 5 & $a_{51}$ & $a_{52}$ & $a_{53}$ & $a_{54}$ \\
        \end{tabular} \\
        (i) \\
        $\begin{bmatrix}
            a_{11} & a_{12} & a_{13} & a_{14} \\
            a_{21} & a_{22} & a_{23} & a_{24} \\
            a_{31} & a_{32} & a_{33} & a_{34} \\
            a_{41} & a_{42} & a_{43} & a_{44} \\
            a_{51} & a_{52} & a_{53} & a_{54}
        \end{bmatrix} \cdot
        \begin{bmatrix}
            t_1 \\
            t_2 \\
            t_3 \\
            t_4 
        \end{bmatrix} = 
        \begin{bmatrix}
            em_1 \\
            em_2 \\
            em_3 \\
            em_4 \\
            em_5
        \end{bmatrix}$ \\
        (ii)
    \end{tabular}
    \label{tab:exampleBB&TC}
\end{table}

{\em Linear Least Squares Approximation} is a mathematical method for estimating the true value of variables in the variable space based on a consideration of errors in measurements. In \textit{"Interactive Linear Algebra"} by Dan Margalit and Joseph Rabinoff~\cite{margalit2017interactive}, the best approximation solution to an inconsistent matrix equation $Ax=b$ is defined as follow:
\begin{tcolorbox}[boxsep=1pt,left=4pt,right=4pt,top=2pt,bottom=2pt]
    {\bf Definition: } Let $A$ be an $m \times n$ coefficient matrix and let $b$ be a vector of objective values in ${\rm I\!R}^m$. A least square solution of the matrix equation $Ax = b$ is a vector $\hat{x}$ in ${\rm I\!R}^n$ such that
    \begin{equation}
        dist(b, A\hat{x}) \leq dist(b, Ax)
    \end{equation}
    for all other vector $x$ in ${\rm I\!R}^n$.
\end{tcolorbox}
Here the distance function is the 2-norm of the vector. Then we can rephrase the linear least square approximation problem to the mathematical formula by following the above definition:
\begin{equation}
    \min ||Ax-b||_2^2 + ||x||_2^2
\end{equation}
As we can see, $||Ax-b||_2$ is the distance from the vector $Ax$ to the vector $b$. To avoid the negative value, we take the power two of the distance value and the ultimate goal is to find the minimum distance between the vector $Ax$ and the vector $b$. In some cases, there will have multiple possible optimal solutions for the problem. Thus, we add a 2-norm of variable vector $x$ so that the variables in $x$ are as small as possible.

Finally, in the {\em evaluation block} in Figure~\ref{fig:System Framework}, {\LESS} collects the normalized test execution time and mutation score from selected test cases. Those two evaluation metrics will be introduced in Section~\S\ref{evaluationMetric}. Since our method is stochastic like other optimizers, we repeat the experiment 20 times for the generality. 

Next we will explain how test case minimization problem with effectiveness measurements can be formulated to the above linear least square problem. First of all, as we presented in Section~\S\ref{blackboxData}, the overall score of each effectiveness measurement is the sum of each individual effectiveness measurement score for selected test cases. As shown in Table~\ref{tab:exampleBB&TC}(i), each test case $T_j$ has its own effectiveness measurement score $a_{ij}$ where $i$ represents the $i$th effectiveness measurement. The overall score of $i$th effectiveness measurement will be calculated by $\sum_{j=1}^4 a_{ij} \times t_j$ where $t_j$ will be the binary variable which represents whether test case $j$ is selected or not. Overall there will have $i$ equations since the number of effectiveness measurements is $i$ (In our study, $i$ = 5). The overall equation systems can be rewritten to the inconsistent matrix equation $Ax=b$ where $A$ has elements $a_{ij}$, $x$ is the vector which contains tests minimization outcome, and $b$ represents the final scores of $i$ effectiveness measurements (The matrix representation is shown in Table~\ref{tab:exampleBB&TC}). At this point, the test case minimization problem is reshaped to the linear least squares approximation problem and the optimal test minimization outcome will be the $x$ which minimize $min ||Ax-b||_2^2$.

However, there is an issue that since there are more tests than the number of effectiveness measurements, solving the linear least square problem can result multiple possible solutions. To find the best one among these possible solutions, we add a term $||x||_2^2$ after the least square formulation which minimizes the overall elements in the vector $x$. Please note that the test case minimization problem is trying to find {\em minimal} tests, so the extra term can make sure the final optimal solution has minimal tests in it.

In the formulation $Ax = b$, where $x$ is an outcome vector of test minimization, we want to predict the value (0/1) for each entry of $x$. In our linear least square formulation, the final outcome of $x$ is a vector of float numbers (ranged from 0 to 1 by controller) which indicates lower effect to the final score with coefficient $\rightarrow$ higher effect to the final score with coefficient from 0 $\rightarrow$ 1. Since test selection outcomes can only have 0 (discard that test) and 1 (select that test), we use the threshold of 0.5 to indicate higher probability or lower probability. A value $<$ 0.5 means higher chance to be 0 and a value $>$ 0.5 means higher chance to be 1. For each representative candidate found by continuous domination, {\LESS} finds the test selection which can get the closest score to that candidate. Although there exists delta between original ideal scores and truth scores generated by optimized results because of the approximation procedure, our results show that least square approximation can find adequate test cases which perform as well or better as the previous state-of-the-art. Our implementation of above step uses {\bf scipy.optimize}, a python library, and uses the function called {\bf lsq\_linear}, which solves the above problem by using either dense QR decomposition technique or Singular Value Decomposition technique.

\begin{table}[!t]
    \centering
    \small
    \caption{Summary of number of I/O signals, number of test cases, and number of mutants in six case studies, due to space situation, in the below table, TT means ``Two Tanks'' project and ACE means ``AC Engine'' project.}
    \begin{tabular}{c||c|c|c|c|c|c}
        Project & TT & CW & ACE & EMB & CC & Tiny \\
        \hline
        \# Inputs & 11 & 15 & 4 & 1 & 5 & 3 \\
        \hline
        \# Outputs & 7 & 4 & 1 & 1 & 2 & 1 \\
        \hline
        \# Test Case & 150 & 133 & 120 & 150 & 150 & 150 \\
        \hline
        \# Mutants & 6 & 96 & 12 & 18 & 20 & 12
    \end{tabular}
    \label{tab:datasummary}
\end{table}

\section{Experimental Setup}\label{experiment}

\subsection{Case Studies}\label{caseStudy}
One challenge with research in this area is that public domain case studies are hard to find (which this has implications for reproducibility of results). After  scouring the  literature on cyber-physical testing we find that most of the studies utilized 2-4 models for their evaluation. 
After collecting all the public models we could find, we find six public Cyber-physical models to evaluate our proposed approach.

For each of those models,
we implemented the test cases and mutants generated by Arrieta et al.'s study\footnote{\url{https://github.com/aitorarrietamarcos/IST2019Paper}}. The summary of number of initial test cases and number of mutants are shown in Table~\ref{tab:datasummary}. In that table: Two Tanks project is a model that simulate the incoming and outgoing flows of the tanks~\cite{menghi2019generating}; CW project is a model that simulate the electrics and mechanics of four car windows~\cite{arrieta2019pareto}; AC Engine project is a model that simulate some safety functionalities in the AC engine~\cite{arrieta2019search, arrieta2017employing}; EMB project simulates the software model controller which includes a continuous PID controller and a discrete state machine~\cite{matinnejad2017automated}; CC is a cruise controller system and Tiny is a simple physical model~\cite{arrieta2019pareto}.

At first glance, the case studies in Table~\ref{tab:datasummary} may appear to contain very small test cases. But appearances can be deceiving; e.g. the number of input signals is a poor measure of the internal complexity of a cyber-physical system. As shown in our {\bf RQ1} results, the systems of  Table~\ref{tab:datasummary} are so complex that, for the purposes of test suite minimization, they defeated state-of-the-art optimizers (NSGA-III and MOEA/D).

Another unusual phenomenon that may cause confusion here is that some case studies have more inputs and outputs but contain less number of mutants than those case studies which has less inputs and outputs (e.g. TT and EMB). Here we need to declare that such phenomenon is normal in Simulink models. The reason is that the complexity of a Simulink model does not depend on the number of inputs and outputs, instead it depends on the complexity of blocks and operations (as we stated in Section~\S\ref{mutation}). Hence, a Simulink model with more inputs and outputs may have less mutants than others due to the model complexity and the number of available operations that can be mutated.

\subsection{Performance Criteria}\label{evaluationMetric}
To evaluate the selected test cases, we use two evaluation metrics from
prior work~\cite{arrieta2019pareto}. 
In summary, a good test case selection approach can both (a) minimize the test execution time and (b) maximize the mutant detection score.

To measure these, we use two  evaluation metrics:  (a) normalized test execution time and (b) mutant detection score. Previous study used these two metrics to calculate the hypervolumne indicator and average weighted sum of mutation score and normalized test execution time~\cite{arrieta2019pareto}, while in our study, we directly compare the performance of algorithms in these two metrics.

{\bf Normalized test execution time} (NTET-): Our goal for selecting test cases from the initial test suite is to speed up the testing process. Thus, test execution time is a very important indicator which can indicate whether selected test cases can significantly reduce the cost of testing. In this study we want to \underline{\em minimize} this value since, as discussed in our introduction, the whole point of this paper is to reduce the time required for testing cyber-physical systems.

{\bf Mutant detection score} (MS+): Given a fixed set of mutants $M_i$
(that conform to  Table~\ref{tab:fault_pattern}), 
and a  test case generation method (e.g. MC/DC coverage, random, other) that has generated a set of tests $T_j$,
and  different  test selection methods can be scored
according to what percent  of $T_j$ have different outputs in the mutants and the original code.
In this way, we say that test selection method1 is better than method2 if that MS percentage is higher
in method1.
Note that another goal (reduction in the overall number of tests) is captured implicitly by the above (TET-) score.

\subsection{Statistical Analysis}\label{statisticalAnalysis}
In our study, we record the value of above two evaluation metrics in 20 repeats. To compare the total performance of different algorithms, we implement A Scott-Knott analysis~\cite{mittas2012ranking}. The Scott-Knott analysis can sort the candidates by their values, and assign candidates to different ranks if the values of candidate at position $i$ is significantly different (by more than a small effect size) to the values of candidate at position $i-1$~\cite{ling2021different}.

More precisely, Scott-Knott sorts the candidates by their median scores
(and in our study, the candidates are the test case selection approaches). Scott-Knott method will split the sorted candidates into two sub-lists which maximize the expected value of differences in the observed performances before and after division~\cite{emblem}. After that, Scott-Knott will declare the one of the split as the best split. The best split should maximize the difference $E(\Delta)$ in the expected mean value before and after the split~\cite{xia2018hyperparameter, 9463120}:
\begin{equation}
    E(\Delta) = \frac{|l_1|}{|l|}abs(\overline{l_1} - \overline{l})^2 + \frac{|l_2|}{|l|}abs(\overline{l_2} - \overline{l})^2
\end{equation}
where $|l|$, $|l_1|$, and $|l_2|$ are size of list $l$, $l_1$, and $l_2$. $\overline{l}$, $\overline{l_1}$, and $\overline{l_2}$ are mean value of list $l$, $l_1$, and $l_2$.

After the best split, Scott-Knott then implements some statistical hypothesis tests to check the division. If two items $d_1$ and $d_2$ after division differ significantly by applying hypothesis test $H$, then such division is defined as a "useful" division. Scott-Knott will run recursively on each half of the best division until no division can be made. In our study, we use cliff's delta non-parametric effect size measure as the hypothesis test. Cliff's delta quantifies the number of difference between two lists of observations beyond p-values interpolation~\cite{xia2018hyperparameter}. The division passes the hypothesis test if it is not a "small" effect ($Delta \geq$ 0.147). The cliff's delta non-parametric effect size test explores two list $A$ and $B$ with size $|A|$ and $|B|$:
\begin{equation}
    Delta = \frac{\sum\limits_{x \in A} \sum\limits_{y \in B} \left\{ \begin{array}{l}
                    +1, \mbox{   if $x > y$}\\
                    -1, \mbox{   if $x < y$}\\
                    0,  \mbox{   if $x = y$}
                \end{array} \right.}{|A||B|}
\end{equation}
Cliff's delta estimates the probability that a value in the list $A$ is greater than a value in the list $B$, minus the reverse probability~\cite{macbeth2011cliff} in the above formula. This hypothesis test and its effect size is supported by Hess and Kromery~\cite{hess2004robust}.

\section{Results}\label{result}

This section is divided into two parts. First, in our {\em Baseline} section, we reproduce past work which shows that test case selection for multi-goal cyber-physical systems is a hard problem. Second, we explore the {\em research questions} defined in our introduction.

\begin{table}[!t]
    \scriptsize
    \centering
    \caption{Replication results of state-of-the-art approach, as well as the off-the-shelf optimizers. MiL \revised{(Model-in-the-Loop)} simulation TET represents the simulation test cases execution time. The metric with "-" means less is better while "+" means more is better. The \colorbox{gray!25}{light gray} cell in each project means that approach wins others significantly (as computed by the statistical method in~\S\ref{statisticalAnalysis}). Moreover, the \colorbox{gray!60}{dark gray} on column \textbf{runtime}(s) marks the significant longer runtime for state-of-the-art NSGA-II method.}
    \begin{tabular}{c|c||c|c||c|c|c}
        \multirow{2}{*}{Project} & \multirow{2}{*}{Approach} & \multicolumn{2}{c||}{MiL TET (s)} & \multirow{2}{*}{NTET-} & \multirow{2}{*}{MS+} & \multirow{2}{*}{runtime} \\
         & & Before & After & & \\
        \hline
        \multirow{3}{*}{Twotanks} & NSGA-II & \multirow{3}{*}{30215} & 9064.5 & \cellcolor{gray!25} 0.30 & \cellcolor{gray!25} 1.00 & \cellcolor{gray!60}11964.6 \\
         & NSGA-III & & 14805 & 0.49 & \cellcolor{gray!25} 1.00 & 2484.3 \\
         & MOEA/D & & 16316 & 0.54 & \cellcolor{gray!25} 1.00 & 1296.2 \\
        \hline
        \multirow{3}{*}{CW} & NSGA-II & \multirow{3}{*}{3913.8} & 1526.3 & \cellcolor{gray!25} 0.39 & \cellcolor{gray!25} 0.99 & \cellcolor{gray!60}15409.9 \\
         & NSGA-III & & 2387.4 & 0.61 & \cellcolor{gray!25} 0.98 & 2185.0\\
         & MOEA/D & & 2661.4 & 0.68 & \cellcolor{gray!25} 0.99 & 1205.2 \\
        \hline
        \multirow{3}{*}{ACEngine} & NSGA-II & \multirow{3}{*}{5530} & 2101.4 & \cellcolor{gray!25} 0.38 & \cellcolor{gray!25} 0.73 & \cellcolor{gray!60} 14042.1 \\
         & NSGA-III & & 3373.3 & 0.61 & 0.72 & 1895.7 \\
         & MOEA/D & & 3594.6 & 0.65 & \cellcolor{gray!25} 0.73 & 1221.7 \\
        \hline
        \multirow{3}{*}{EMB} & NSGA-II & \multirow{3}{*}{4830} & 1787.1 & \cellcolor{gray!25} 0.37 & \cellcolor{gray!25} 1.00 & \cellcolor{gray!60} 12585.6 \\
         & NSGA-III & & 2608.2 & 0.54 & \cellcolor{gray!25} 1.00 & 2019.6 \\
         & MOEA/D & & 3042.9 & 0.63 & \cellcolor{gray!25} 1.00 & 1300.9 \\
        \hline
        \multirow{3}{*}{CC} & NSGA-II & \multirow{3}{*}{3142} & 911.1 & \cellcolor{gray!25} 0.29 & \cellcolor{gray!25} 0.99 & \cellcolor{gray!60} 12522.8 \\
         & NSGA-III & & 1508.1 & 0.48 & \cellcolor{gray!25} 1.00 & 1922.7 \\
         & MOEA/D & & 1728.1 & 0.55 & \cellcolor{gray!25} 1.00 & 1306.8 \\
        \hline
        \multirow{3}{*}{Tiny} & NSGA-II & \multirow{3}{*}{811} & 291.9 & \cellcolor{gray!25} 0.36 & 0.98 & \cellcolor{gray!60} 15696.0 \\
         & NSGA-III & & 413.6 & 0.51 & \cellcolor{gray!25} 1.00 & 2572.3 \\
         & MOEA/D & & 486.6 & 0.60 & \cellcolor{gray!25} 1.00 & 1408.5
    \end{tabular}
    \label{tab:rq1}
\end{table}

{\bf Baseline Task:}
To begin,  
we demonstrate  that, using
prior methods, multi-goal cyber-physical test generation is hard task.
  Arrieta et al.'s experiments~\cite{arrieta2019pareto} are replicated in  Table~\ref{tab:rq1} (using 20 repeats, with 20 different random number seeds). In that table, two algorithms differ significantly if they separate in different ranks in the Scott-Knott analysis.  

As seen in Table~\ref{tab:rq1}, we can find the approach with NSGA-II that Arrieta et al. proposed~\cite{arrieta2019pareto} performs better than other multi-goal optimizers (MOEA/D and NSGA-III). 
Specifically, in all six case studies, NSGA-II can find the most minimal test suite among 3 off-the-shelf search algorithms while the minimized test suite from these 3 algorithms can achieve similar mutation scores. This means the feedback loop of executing test suite minimized by NSGA-II is much shorter than those minimized by NSGA-III and MOEA/D (We will discuss the indicator of why NSGA-III and MOEA/D may failed in this task in discussion section later).
However, NSGA-II has limitations on the number of optimized goals as we discussed in~\S\ref{background}. Specifically, NSGA-II can only handle two or three goals which is verified by Panichella et al.~\cite{panichella2017automated}. Hence, the state-of-the-art approach has to be re-run multiple times to explore all pairs of five goals, which requires much longer feedback loop during the test case minimization as shown in the last column in Table~\ref{tab:rq1}. As shown below, we can achieve similar or better results 80-360 times faster. Moreover, as shown in MiL TET column of Table~\ref{tab:rq1}, with shorter execution time on test cases (e.g. Tiny), test case selection can:
\begin{itemize}[leftmargin=4mm]
\item
Reduce the test execution time of the small projects by at least several minutes;
\item
And with larger systems  (e.g. Twotanks), it can reduce the test execution time by several hours.
\end{itemize}
Note that the more complex use cases for a system, the more impressive are our time savings.  
Considering the problem of testing larger tasks (e.g. drone simulation, elevator simulation).
In the real world, a single test case can take several hours to few days in a simulator. For those larger tasks, it is vital that we find the best representative test cases to reduce the overall testing budget. Also, all CPSs need to be tested in the real time simulation, which can be multiple times lengthy than MiL testing. Therefore, it is very important and necessary to reduce the number of test cases in the MiL.
Hence, we say:
\begin{tcolorbox}[boxsep=1pt,left=4pt,right=4pt,top=2pt,bottom=2pt]
    Test case selection for multi-goal cyber-physical models is a {\it hard problem} that cannot be solved by merely applying, off-the-shelf, the recent optimizer technology (e.g. NSGA-III and MOEA/D). Moreover, the state-of-the-art approach can address this problem by utilizing NSGA-II in their algorithm, but needs extensive CPU execution before finding the best choice. To summarize that, developing a multi-goals optimizer with fast feedback loop is a hard problem in test case selection for cyber-physical systems.
\end{tcolorbox}

{\bf RQ1: How effective is {\LESS}?}
To answer RQ1, we compare scores of NTET- ({\em normalized test execution time}) and MS+ ({\em mutant detection score}) between the prior state-of-the-art and {\LESS}. It is important to have higher performance on these two evaluation metrics since they directly indicate whether a test case selection is good or not. 

\begin{table}[t!]
    \centering
    \scriptsize
    \caption{RQ1 results: Scores of two evaluation metrics calculated by selected test cases. All entries are reported the median score of 20 repeats. In the title row, the metric with ``-'' means less is better while ``+'' means more is better. The \colorbox{gray!25}{light gray} cells mark the winning approach (as computed by the statistical method in~\S\ref{statisticalAnalysis}) in that metric. Last column counts the number of wins for each approach.}
    \begin{tabular}{c|c||c|c||c|c||c}
        \multirow{2}{*}{Project} & \multirow{2}{*}{Approach} & \multicolumn{2}{c||}{MiL TET (s)} & \multirow{2}{*}{NTET-} & \multirow{2}{*}{MS+} & \multirow{2}{*}{Wins} \\
        & & Before & After & & & \\
        \hline
        \multirow{2}{*}{Twotanks} & NSGA-II & \multirow{2}{*}{30215} & 9064.5 & \cellcolor{gray!25} 0.30 & \cellcolor{gray!25} 1.00 & {\bf 2} \\
                                  & \LESS & & 9064.5 & \cellcolor{gray!25} 0.30 & \cellcolor{gray!25} 1.00 & {\bf 2} \\
        \hline
        \multirow{2}{*}{CW} & NSGA-II & \multirow{2}{*}{3913.8} & 1526.3 & 0.39 & \cellcolor{gray!25} 0.98 & {\bf 1} \\
                            & \LESS & & 1408.9 & \cellcolor{gray!25} 0.36 & 0.95 & {\bf 1}\\
        \hline
        \multirow{2}{*}{ACEngine} & NSGA-II & \multirow{2}{*}{5530} & 2010.4 & 0.39 & \cellcolor{gray!25} 0.72 & 1 \\
                                  & \LESS & & 1659 & \cellcolor{gray!25} 0.30 & \cellcolor{gray!25} 0.72 & {\bf 2} \\
        \hline
        \multirow{2}{*}{EMB} & NSGA-II & \multirow{2}{*}{4830} & 1787.1 & 0.37 & \cellcolor{gray!25} 1.00 & 1 \\
                             & \LESS & & 1690.5 & \cellcolor{gray!25} 0.35 & \cellcolor{gray!25} 1.00 & {\bf 2} \\
        \hline
        \multirow{2}{*}{CC} & NSGA-II & \multirow{2}{*}{3142} & 911.1 & \cellcolor{gray!25} 0.29 & \cellcolor{gray!25} 0.99 & {\bf 2} \\
                                  & \LESS & & 911.1 & \cellcolor{gray!25} 0.29 & \cellcolor{gray!25} 1.00 & {\bf 2} \\
        \hline
        \multirow{2}{*}{Tiny} & NSGA-II & \multirow{2}{*}{811} & 291.9 & 0.36 & 0.98 & 0 \\
                             & \LESS & & 251.4 & \cellcolor{gray!25} 0.31 & \cellcolor{gray!25} 1.00 & {\bf 2} \\
    \end{tabular}
    \label{tab:rq2}
\end{table}

Table~\ref{tab:rq2} shows our simulation results (note that TET - test execution time \& MS - Mutation Score). For each method in each project, we repeat experiment 20 times and calculate the value of two evaluation metrics for each repeat. To obtain the final conclusion, we implement Scott-Knott statistical method to check if our approach significantly differ to state-of-the-art approach in each metric. The \colorbox{gray!25}{light gray} cells mark the winning method (in the first rank) resulted from the Scott-Knott test. Moreover, we count the number of higher performance in these two evaluation metrics for each algorithm and record the number of wins in the last column.

As seen in Table~\ref{tab:rq2}, {\LESS} gets better performance in both two evaluation metrics in three out of six projects (ACEngine, EMB, \& Tiny). Moreover, in Twotanks and CC projects, {\LESS} and state-of-the-art method achieve the same performance (both two evaluation metrics are tied in the first rank). In the CW project, {\LESS} has higher performance in minimizing test execution time when state-of-the-art method gets a little bit higher mutation score than {\LESS}. Taking above comparisons together, we can conclude that test cases selected by {\LESS} can achieve similar or better performance in minimizing test execution time and detecting more mutants in all projects.

By summarizing above findings, we answer {\bf RQ1} as follow:
\begin{tcolorbox}[boxsep=1pt,left=4pt,right=4pt,top=2pt,bottom=2pt]
    In {\bf all} projects, comparing to the state-of-the-art, {\LESS} can get {\it similar or better} performance on minimizing the execution time of selected test cases while {\it keeping} to detect most of the mutants. 
\end{tcolorbox}

{\bf RQ2: How fast is {\LESS}}
To answer RQ2, we count the execution time for both algorithms during the experiment. To make comparison fair enough, we run both two algorithms on the same 64-bit Windows 10 machine with a 4.2 GHz 8-core Intel Core i7 processor and 16 GB of RAM. Moreover, when running experiments, we make sure no huge process is starting or ending in our machine. 

Table~\ref{tab:rq3} shows the recorded runtime for each project. For each method, we repeat experiments 20 times and record the total runtime. The \colorbox{gray!25}{light gray} cells mark the fastest approach.

\begin{table}[t!]
    \centering
    \footnotesize
    \caption{RQ2 results: Runtime comparison for our proposed {\LESS} and other optimizers (note NSGA-II-5 means NSGA-II on 5 goals) on totally 20 repeats, as well as the min/max time and standard deviation through 20 repeats. The \colorbox{gray!25}{light gray} cell marked the fastest approach in each project. The last column marks how many \textbf{times faster} our proposed approach compare to other optimizers.}
    \begin{adjustbox}{max width=.48\textwidth}
    \begin{tabular}{c|c||c|c|c|c||c}
        \multirow{2}{*}{Project} & \multirow{2}{*}{Approach} & \multicolumn{4}{c||}{Stats (s)} & \multirow{2}{*}{Slower} \\
         & & min & max & std & overall & \\
        \hline
        \multirow{5}{*}{Twotanks} & MOEA/D & 62.7 & 67.9 & 1.6 & 1296.2 & 9 \\
         & NSGA-III & 103.2 & 132.7 & 8.6 & 2484.3 & 18 \\
         & NSGA-II-5 & 119.3 & 163.4 & 14.6 & 2913.4 & 21 \\
         & NSGA-II & 721.6 & 815.3 & 23.3 & 11964.6 & 83 \\
         & \LESS & 6.9 & 7.5 & 0.2 & \cellcolor{gray!25} 144.7 & - \\
        \hline
        \multirow{5}{*}{CW} & MOEA/D & 59.3 & 63.5 & 1.2 & 1205.2 & 29 \\
         & NSGA-III & 49.6 & 142.6 & 28.6 & 2185.0 & 52 \\
         & NSGA-II-5 & 97.6 & 131.9 & 11.7 & 3001.0 & 71\\ 
         & NSGA-II & 666.7 & 808.9 & 19.9 & 15409.9 & 362 \\
         & \LESS & 2.0 & 2.2 & 0.1 & \cellcolor{gray!25} 42.6 & - \\
        \hline
        \multirow{5}{*}{ACEngine} & MOEA/D & 55.3 & 57.86 & 0.98 & 1221.7 & 16 \\
         & NSGA-III & 92.7 & 96.1 & 1.08 & 1895.7 & 25 \\
         & NSGA-II-5 & 109.8 & 147.1 & 12.50 & 2613.7 & 34 \\ 
         & NSGA-II & 665.1 & 733.9 & 18.0 & 14042.1 & 179 \\
         & \LESS & 3.5 & 4.3 & 0.2 & \cellcolor{gray!25} 78.6 & - \\
        \hline
        \multirow{5}{*}{EMB} & MOEA/D & 62.7 & 68.4 & 1.4 & 1300.9 & 33 \\
         & NSGA-III & 96.2 & 106.3 & 3.1 & 2019.6 & 52 \\
         & NSGA-II-5 & 116.6 & 169.6 & 15.7 & 3109.8 & 79\\ 
         & NSGA-II & 545.7 & 711.8 & 14.5 & 12585.6 & 319 \\
         & \LESS & 1.9 & 2.4 & 0.1 & \cellcolor{gray!25} 39.5 & - \\
        \hline
        \multirow{5}{*}{CC} & MOEA/D & 61.8 & 68.5 & 2.3 & 1306.8 & 10 \\
         & NSGA-III & 92.3 & 100.0 & 2.3 & 1922.7 & 14 \\
         & NSGA-II-5 & 129.1 & 170.0 & 10.7 & 3175.8 & 23 \\ 
         & NSGA-II & 515.1 & 675.3 & 16.8 & 12522.8 & 89 \\
         & \LESS & 6.4 & 8.7 & 0.6 & \cellcolor{gray!25} 140.9 & - \\
        \hline
        \multirow{5}{*}{Tiny} & MOEA/D & 62.7 & 77.1 & 4.7 & 1408.5 & 19 \\
         & NSGA-III & 106.3 & 137.1 & 10.5 & 2572.3 & 35 \\
         & NSGA-II-5 & 112.1 & 171.6 & 24.0 & 2741.3 & 37 \\ 
         & NSGA-II & 541.6 & 857.9 & 117.3 & 15696.0 & 210 \\
         & \LESS & 2.6 & 4.7 & 0.7 & \cellcolor{gray!25} 74.6 & - \\
    \end{tabular}
    \end{adjustbox}
    \label{tab:rq3}
\end{table}

\begin{table*}[!t]
    \centering
    \scriptsize
    \caption{The sanity check results of 5 different approaches for each case study. In the right-hand-side of the header row, the entry with ``-'' at the end means minimization goal and the entry with ``+'' at the end means maximization goal.}
    \begin{tabular}{c|c|c|c|c|c|c|c|c|c} 
        \multirow{2}{*}{Project} & \multirow{2}{*}{Approach} & \multirow{2}{*}{Best Combination} & \multicolumn{3}{c|}{Efficacy goals} & \multicolumn{4}{c}{Optimization goals} \\ 
         & & & MS+ & TET- & Slower- & Discontinuity+ & Infinity+ & Instability+ & MinMax+ \\
        \hline
        \multirow{5}{*}{Twotanks} & MOEA/D & - & 1.00 & 0.54 & 9 & 0.81 & 0.92 &  0.83 &  0.92 \\
         & NSGA-III & - & 1.00 & 0.49 & 18 & 0.70 & 0.81 & 0.75 &  0.81 \\
         & NSGA-II & - & 1.00 & 0.35 & 21 & 0.58 & 0.67 & 0.64 & 0.67 \\
         & NSGA-II & Time, Infinite, Minmax & 1.00 & 0.30 & 83 & 0.54 & 0.65 & 0.55 & 0.65 \\
         & \LESS & - & 1.00 & 0.30 & 1 & 0.56 & 0.67 & 0.56 & 0.67 \\
        \hline
        \multirow{5}{*}{CW} & MOEA/D & - & 0.99 & 0.68 & 29 & 0.82 &  0.79 &  0.84 & 0.79 \\
         & NSGA-III & - & 0.98 & 0.61 & 52 & 0.74 &  0.67 &  0.76 &  0.67 \\
         & NSGA-II & - & 0.98 & 0.46 & 71 & 0.60 & 0.57 & 0.60 & 0.57 \\
         & NSGA-II & Time, Instability & 0.98 & 0.39 & 362 & 0.55 &
        0.34 & 0.64 & 0.34 \\
         & \LESS & - & 0.95 & 0.36 & 1 & 0.50 & 0.58 & 0.44 & 0.58 \\
        \hline
        \multirow{5}{*}{ACEngine} & MOEA/D & - & 0.73 & 0.65 & 16 & 0.84 & 0.80 &  0.78 &  0.80 \\
         & NSGA-III & - & 0.72 & 0.61 & 25 & 0.77 &  0.74 & 0.72 &  0.74 \\
         & NSGA-II & - & 0.72 & 0.40 & 34 & 0.56 & 0.53 & 0.51 & 0.53 \\
         & NSGA-II & Time, Discontinuity, Instability & 0.72 & 0.39 & 179 & 0.53 & 0.49 & 0.49 & 0.49 \\
         & \LESS & - & 0.72 & 0.30 & 1 & 0.47 & 0.44 & 0.41 & 0.44 \\
        \hline
        \multirow{5}{*}{EMB} & MOEA/D & - & 1.00 & 0.63 & 33 & 0.83 &  0.83 &  0.84 & 0.83 \\
         & NSGA-III & - & 1.00 & 0.54 & 52 &  0.70 & 0.70 &  0.70 &  0.70 \\
         & NSGA-II & - & 1.00 & 0.42 & 79 & 0.61 & 0.61 & 0.59 & 0.61 \\
         & NSGA-II & Time, Instability & 1.00 & 0.37 & 319 & 0.50 & 0.49 & 0.57 & 0.50 \\
         & \LESS & - & 1.00 & 0.35 & 1 & 0.61 & 0.61 & 0.48 & 0.61 \\
        \hline
        \multirow{5}{*}{CC} & MOEA/D & - & 1.00 & 0.55 & 10 & 0.87 &  0.86 &  0.88 &  0.86 \\
         & NSGA-III & - & 1.00 & 0.48 & 14 & 0.77 & 0.74 &  0.76 &  0.74 \\
         & NSGA-II & - & 1.00 & 0.35 & 23 & 0.64 & 0.58 & 0.61 & 0.58 \\
         & NSGA-II & Time, Discontinuity & 0.99 & 0.29 & 89 & 0.62 & 0.43 & 0.50 & 0.43 \\
         & \LESS & - & 1.00 & 0.29 & 1 & 0.56 & 0.56 & 0.57 & 0.56 \\
        \hline
        \multirow{5}{*}{Tiny} & MOEA/D & - & 1.00 & 0.60 & 19 & 0.83 &  0.83 &  0.84 & 0.83 \\
         & NSGA-III & - & 1.00 & 0.51 & 35 & 0.72 & 0.71 & 0.73 & 0.71 \\
         & NSGA-II & - & 0.99 & 0.40 & 37 & 0.63 & 0.62 & 0.66 & 0.62 \\
         & NSGA-II & Time, Instability & 0.98 & 0.36 & 210 & 0.57 & 0.56 & 0.64 & 0.56 \\
         & \LESS & - & 1.00 & 0.31 & 1 & 0.58 & 0.57 & 0.57 & 0.57
    \end{tabular}
    \label{tab:sanitycheck}
\end{table*}

As seen in Table~\ref{tab:rq3}, in all six projects, {\LESS} runs significantly faster (80-360 times faster) than the previous state-of-the-art method from 20 repeats. Moreover, the reported min/max time and their standard deviation in each single run shows that all approaches are stable and no outlier included in those 20 repeats. By analyzing our proposed algorithm and state-of-the-art approach, we find previous approach implemented NSGA-II as their multi-objective optimization, which designed for 2 or 3 objectives~\cite{panichella2017automated}. To handle this issue, Arrieta et al. group objectives into 21 different combinations with two or three objectives in each combination, and select one of the best combinations by repeating their approach in those 21 groups~\cite{arrieta2019pareto}. However, in our approach, we just need continuous domination to find ``ideal goals'' and approximate corresponding test cases inversely.

Moreover, comparing {\LESS} to other off-the-shelf optimizers (e.g. NSGA-III, MOEA/D, and NSGA-II on 5 goals), we can find {\LESS} can still run significantly faster than these optimizers. Moreover, in the next section, we will discuss that {\LESS} can solve the issues which cannot be solved by these optimizers.

By summarizing above findings, we answer {\bf RQ2} as follow:
\begin{tcolorbox}[boxsep=1pt,left=4pt,right=4pt,top=2pt,bottom=2pt]
    In {\bf both six} projects, {\LESS} run significantly faster (80-360 times) than the previous method. In other words, {\LESS} is far more efficient than state-of-the-art approach.
\end{tcolorbox}

Even though our current empirical results can only boost a speed of (up to) 300 times faster, we can make a theoretical case that, if the number of goals increases, our technique would be even more comparatively faster (please note that in the following counts, test execution time is the metric that must be in every combination):
\begin{itemize}[leftmargin=6mm]
    \item {\bf 5} goals will result {\bf 10} different combinations of 2 or 3 objectives with text execution time must be included.
    \item {\bf 7} goals will result {\bf 21} different combinations of 2 or 3 objectives with text execution time must be included.
    \item {\bf 9} goals will result {\bf 36} different combinations of 2 or 3 objectives with text execution time must be included.
    \item The above pattern shows that with more goals being utilized, the number of repeats for state-of-the-art NSGA-II approach increases exponentially. However, our approach can handle multiple goals simultaneously in one time. This can show the efficiency of our approach.
\end{itemize}

\begin{figure*}
    \centering
    \begin{subfigure}[b]{0.245\textwidth}
         \centering
         \includegraphics[width=\textwidth]{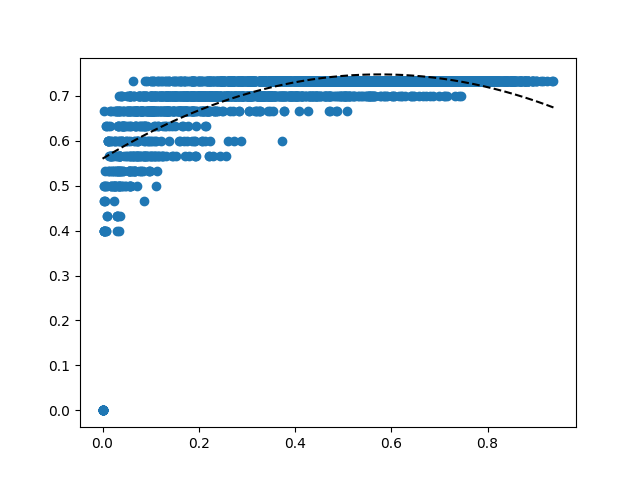}
         \caption{Discontinuity}
         \label{fig:discontinuity}
    \end{subfigure}
    \begin{subfigure}[b]{0.245\textwidth}
         \centering
         \includegraphics[width=\textwidth]{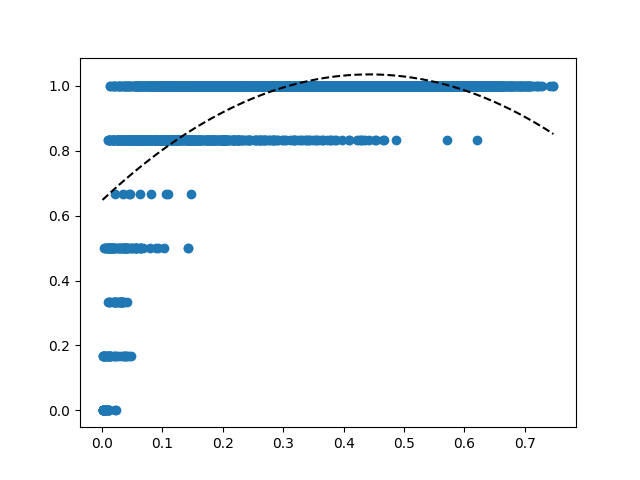}
         \caption{Instability}
         \label{fig:instability}
    \end{subfigure}
    \begin{subfigure}[b]{0.245\textwidth}
         \centering
         \includegraphics[width=\textwidth]{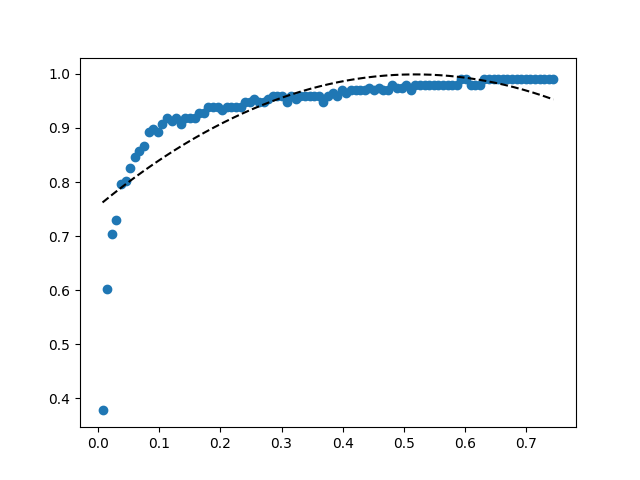}
         \caption{Growth to infinity}
         \label{fig:infinity}
    \end{subfigure}
    \begin{subfigure}[b]{0.245\textwidth}
         \centering
         \includegraphics[width=\textwidth]{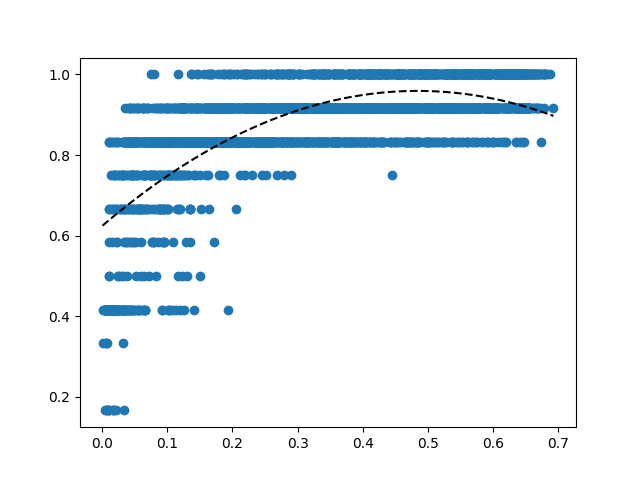}
         \caption{Minmax difference}
         \label{fig:minmax}
    \end{subfigure}
    \caption{Diminishing returns in the $y$-axis as the optimization goals increase along the $x$-axis ($y$-axis shows the mutant detection score). This figure shows there is limited improvement in the $y$ value above an threshold of $x=0.45$ to $x=0.5$. For space reason, this figure shows a small sample of these charts (full figures are in~\url{https://github.com/ai-se/DoLesS}).}
    \label{fig:discussion2}
\end{figure*}

{\bf RQ3: Does maximizing for goal A (mutation effectiveness) mean compromising on optimization goals B?} To find an answer for this research question, we run the sanity check experiment on 5 different approaches, and record the five optimization goals for the final selected test cases. Table~\ref{tab:sanitycheck} shows the overall statistics on those approaches. As we can see, {\LESS} and prior state-of-the-art (NSGA-II) achieve better results on mutation effectiveness (from RQ1), but compromise on four out of five optimization goals. Note that in the actual search, the test execution time is also included to balance the search indicator since only maximizing the optimization goals can make the number of selected test cases not optimal. \revised{The consistent finding through all projects is that pursuing optimization goals can hurt the efficacy goals (especially in TET-). This finding motivates the follow up research which will be stated in RQ4.}

At first glance, these RQ3 results seem to tell a negative story about {\LESS}. But it turns out that this first impression is misleading, for two reasons.

{\em Firstly}, we note that {\LESS}’s supposedly-worse results in \revised{Table~\ref{tab:sanitycheck}} only appears in the mutated systems. That is, we only observe {\LESS} being inferior in corrupted mutated code. This is not very informative  (since those results come from optimizing the wrong program).

{\em Secondly}, it turns out that the relationship between mutation effectiveness and optimization has some interesting non-linear effects. For more on that point, we turn to RQ4.

{\bf RQ4: Should mutation testing tools for cyber-physical systems emphasis optimization goals?}
Figure~\ref{fig:discussion2} shows what happens to the mutation score (on the y-axis) as we make improvements on the optimization goals
(on the x-axis). To generate these plots, we randomly select different test cases from the initial test suite with test suite size from 1 to all test cases. Then we calculate their mutation scores and anti-pattern optimization scores.
More specifically, to select different combinations of tests, 50 times are iterated from 1 to size of the test suite for each model/effectiveness score, and those combinations are summarized as 5000 $(x,y)$ pairs:
\begin{itemize}[leftmargin=4mm]
\item 
Where $x$ is the optimization measurements (anti-patterns);
\item 
And $y$ is the mutation score of that test cases combination.
\item 
If many pairs $\{ (x_i, y_1), (x_i, y_2), (x_i, y_3), \cdots, (x_i, y_k)\}$ have the same effectiveness value $x_i$, we replace those pairs with one pair of $x=x_i$ and $y = median(\{y_1, y_2, \cdots, y_k\})$. 
\end{itemize}
After that, we use scatter plot to reflect all those pairs in a single plot, and draw a polynomial fitting curve\footnote{We tried different parametric forms and found polynomial curve is the best curve to reflect our finding.}. Note the non-linear relationship between  
optimization goals and mutation effectiveness:
\begin{itemize}[leftmargin=4mm]
\item
Initially, they are not positively correlated.
\item
But after some turning point (at around $0.45 <= x <= 0.5$) it turns out the most effective way to achieve better mutation scores is {\em not} to push harder on the  optimization goals
\end{itemize}
If this result surprises readers, we note that there is precedence for such a conclusion in the SE literature. At the end of our introduction, we listed prior studies~\cite{ratner2019training,Pornprasit23,wang2019characterizing,nair2017using} which achieved good optimizing results for some goals $A$ where guided by some easy-to-compute values $B$ (even though $B$ was  only a weak indicator for $A$).

So are these optimization goals effective for guiding mutation analysis?
We say ``yes, up to a point". But after around $0.45 <= x <= 0.5$,
the optimization goals have to relax in order to efficiently select better test cases.

For this reason, we say that it is an important feature of
{\LESS} that it does not ``dive too deep'' (as it where) into the optimization information. Rather, {\LESS} finds optimal test cases by averaging over a coarse grained view of the optimization information
(a cluster tree that divides the data into samples of size $\sqrt{N}$).

\section{Discussion}\label{discussion}

\subsection{The Generality of {\LESS}}
In this section we will discuss the generality of our proposed method {\LESS}.

{\LESS} implements the idea from Chen et al.'s study that uses a domination predicate to sort the space of possible goals to a small number of representative examples. Many-objective continuous domination then splits these candidates into two regions where one region contains samples that dominate the samples in another region. After that, it applies an inverted least square approximation approach to find a minimal set of tests. In cyber-physical system test case selection problem that utilizes the effectiveness measurements, least square approximation is considered as an adequate approach to inversely find the minimal test suite from objective space to variable space because the formulation of this problem from the variable space to the objective space can be treated as a linear equation system (See section~\S\ref{less}). Thus, the first possible adequate application of our method in other software engineering domain is the multi-objective optimization problems which the relation of the variable space to the objective space can be re-formulated as a linear equation system. In such problems, {\LESS} can be directly applied with only small modifications on the formulation.
~\\\indent
However, a large proportion of problems in software engineering domain cannot be re-formulated as a linear equation system since the relationship between variable space and objective space is non-linear. In such scenarios, our proposed {\LESS} approach needs to be modified by changing the linear least square approximation to other approaches (e.g. non-linear least square approximation). However, we hypothesis that the overall framework of {\LESS} can still be applied as a fast method in these multi-objective optimization problems.

\section{Threats to Validity}\label{threats_to_validity}
This section discusses issues raised by Feldt et al.~\cite{feldt2010validity}.

{\bf Construct validity:} The construct validity threat mainly exists in the parameter settings of algorithms. For example, in our replication experiment, we use one point crossover with 0.8 crossover probability and bitflip mutation with $1/$(number of variables) mutate probability as prior studies did in order to keep consistent. For another example, in least square approximation, we use 0.5 threshold to indicate whether a test case has large probability to be chose or not. Moreover, we use the default setting of Python least square solver in our algorithm. Changing these parameters can result differences in selecting test cases. Therefore, our observation may differ when different parameters are used. We would consider hyper-parameter tuning~\cite{9463120, Tu18Tuning} in future work to mitigate this threat.

{\bf Conclusion validity:} The conclusion validity threat in this study is related to the random variations of our algorithm and the access to the real faults. To reduce the effect caused by this threat, we repeat all experiments 20 times in the same machine. Moreover, we apply Scott-Knott statistical test to compare if the outcomes of our proposed approach and the previous methods differ significantly. Moreover, the systems we test are all public systems thus no faulty versions can be utilized in our analysis. To mitigate this threat, we use the same mutants which created by Arrieta et al. ~\cite{arrieta2019pareto} based on common system faulty behaviors, which can make mutants more similar to the real faults. We use these mutants to better address the threat that caused by lacking access to the real faults.

{\bf Internal validity:} Internal validity focuses on the correctness of the treatment caused the outcome. In this study, we constraint our simulations to the same data set. Moreover, we evaluate our approach and the previous approach in the same workflow. Another internal validity threat can refer to the mutants generated from the projects. To mitigate this threat, we use the same mutants that Arrieta et al.~\cite{arrieta2019pareto} used in their study, which they have removed duplicated mutants.

{\bf External validity:} External validity concerns the application of our algorithm in other problems. In this study, we generate our conclusion from six real-world Simulink cyber-physical systems. When applying our method into other case studies, these concerns may raised: (a) Six real-world open-source systems may not represent the society of cyber-physical system. However, to the best of our knowledge, these six models are the most commonly used models in previous studies. Moreover, cyber-physical models are hard to obtain thus almost all previous studies only utilize 3 to 4 models (either private or public). Hence, we used all current resources to better address this threat in this study. In the future, more models can be used to test {\LESS} if they become available. (b) {\LESS} may needs modification for those projects which effectiveness measurement data and test cases are not in the linear relationship. For those projects, we would consider non-linear least square approximation as future work to mitigate this threat.

\section{Conclusion \& Future Work}\label{conclusion}
Finding representative test cases from the initial test suite is an important task in simulation based model. Better test case selection methods can not only reduce the test execution time in the future testing in different test level, but can also maintain the same testing performance as usual. In other words, a good test case selection approach can (a) minimize the test execution time and (b) maximize the mutation score.

Previous literature by Arrieta et al.~\cite{arrieta2019pareto} has shown the great success on using NSGA-II as the multi-objective optimization method to select representative test cases. However, their design has a deficiency where they need to evaluate 21 combinations first to select the best subset. Moreover, since NSGA-II is a randomized algorithm, repeats are necessary during the experiment. Therefore, their approach will execute NSGA-II 420 times with 20 repeats.

In this study, we address this deficiency by selecting test cases from all effectiveness measurement metrics. To do that, we use a very fast approach - continuous domination to select representative goals. Moreover, we make a better use of the linear relationship between test cases and goals to find the best test selections correspond to the representative goals (by linear least square approximation). Our experimental results show that our proposed approach {\LESS} can achieve higher mutation efficacy goal $A$ (similar or better results on $A_1$ and 80-360 times faster on $A_2$) comparing to state-of-the-art and other optimizers. Moreover, we dive deep into the relationship between mutation efficacy $A$ and optimization goals $B$, and found non-linear relation between those two goals. More specifically, up until some point, chasing $B$ does improve mutation efficacy $A$, but after some turning point, further pursuit of optimization goals will not improve the mutation effectiveness too much. Hence, we conclude that (a) selecting test cases via mutation is a {\em different goal} to domain-specific optimization goals, and (b) while the optimization goals can be used to guide the mutation analysis, that guidance should be viewed as a weak indicator since it can hurt the mutation efficacy due to focusing too much on the optimization goals.

We conjecture that our method would be a better candidate for scaling to large systems than the method proposed by Arrieta et al.~\cite{arrieta2019pareto}. To see that, consider the following scenario: To successfully perform test case selection on selected cyber-physical case studies, Arrieta et al.'s approach required several hours algorithm execution time.
Now imagine in some higher complexity simulation models (e.g. drone simulation models) with dozens more of test cases in the initial test suite, and these models have more signal processing criteria in I/O signals, both evaluation time and objectives are increased. In such scenario, the execution time of running NSGA-II in all subsets of objectives will exponentially increase as we mentioned at the end of {\bf RQ2}. Moreover, such models (e.g. drone simulation models) require much faster feedback than usual cyber-physical models. Due to the above reasons, the ideal test case selection approach for complex simulation models needs to handle multiple goals (more than 4 goals) in the same time and perform selection in a very short execution time for the fast feedback.

As to further work, apart from extending this exploration of  feedback loop anti-patterns, we conjecture that our methods could be useful for other multi-objective reasoning tasks. Standard practice in this area is to mutate large populations across a Pareto frontier. This has certainly been a fruitful research agenda~\cite{maia2009multi,yoo2010using,panichella2014improving,zheng2016multi}. But perhaps the testing community could reason about more goals, faster, if used our domination methods and least squares methods to ``reason backwards'' from goal space to decision space. Hence, future works can be conducted with
\begin{itemize}[leftmargin=6mm]
    \item Finding more simulation projects which can strength our approach.
    \item Developing more effectiveness measurement metrics which can better indicate representative test cases.
    \item Adjusting our approach based on the testing scenarios of different projects. Moreover, in some cases, a new test case selection approach is needed for those projects.
\end{itemize}

\section*{Acknowledgments}
This research was partially funded by the National Science Foundation under Grant No. 1908762. Any opinions, findings, and conclusions or recommendations expressed in this material are those of the author(s) and do not necessarily reflect the views of the National Science Foundation.
  
\balance
\bibliographystyle{IEEEtran}
\bibliography{main}

\begin{IEEEbiography}[{\includegraphics[width=1.05in,clip,keepaspectratio]{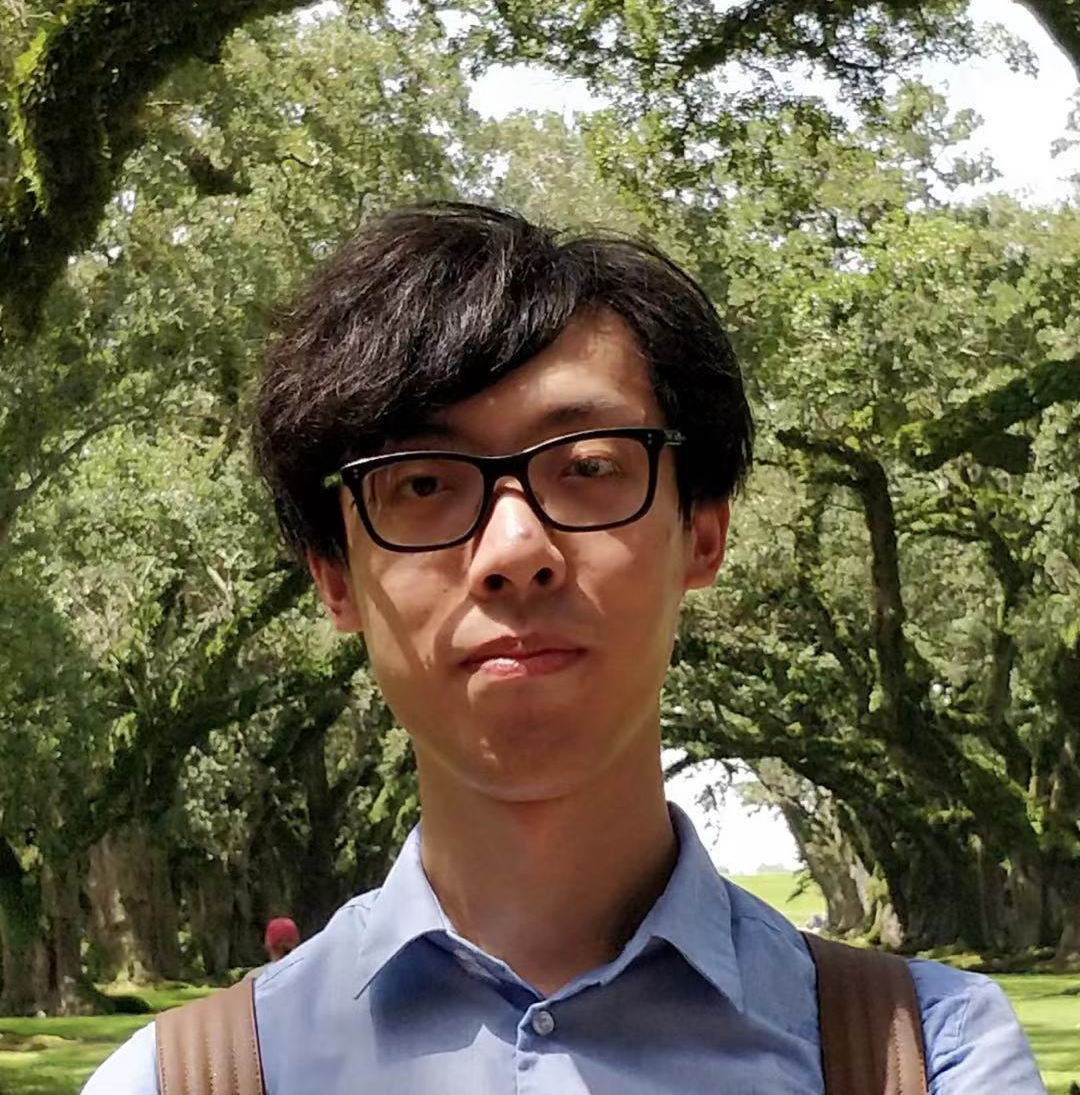}}]{Xiao Ling} is a third-year PhD student in Computer Science at NC State University. His research interests include automated software testing and machine learning for software engineering.
\end{IEEEbiography}

\begin{IEEEbiography}[{\includegraphics[width=1.05in,clip,keepaspectratio]{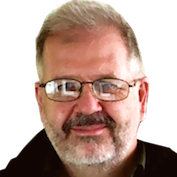}}]{Tim Menzies} (IEEE Fellow, Ph.D. UNSW, 1995)
is a Professor in computer science  at NC State University, USA,  
where he teaches software engineering,
automated software engineering,
and programming languages.
His research interests include software engineering (SE), data mining, artificial intelligence, and search-based SE, open access science. 
For more information,  please visit \url{http://menzies.us}.
\end{IEEEbiography}

\end{document}